\documentclass[journal]{IEEEtran}
\usepackage{graphicx}
\usepackage{epsfig}
\usepackage{amssymb}
\usepackage[numbers]{natbib}
\usepackage{multicol}
\usepackage{multirow}

\begin{document}

%
\title{Memristor MOS Content Addressable Memory (MCAM): Hybrid Architecture for Future High Performance Search Engines}

%
%
%


\author{Kamran~Eshraghian,
        Kyoung-Rok~Cho,~\IEEEmembership{Member,~IEEE,}
        Omid~Kavehei,~\IEEEmembership{Student Member,~IEEE,}
        Soon-Ku~Kang,
        Derek~Abbott,~\IEEEmembership{Fellow,~IEEE,}
        and~Sung-Mo~Steve~Kang,~\IEEEmembership{Fellow,~IEEE}
\thanks{Manuscript received December 31, 2009. This work was supported by grant No.~R33-2008-000-1040-0 from the World Class University (WCU) project of MEST and KOSEF through Chungbuk National University (CBNU).}
\thanks{K.~Eshraghian, K.~R.~Cho, O.~Kavehei, and S.~K.~Kang~are~with the College of Electrical and Information Engineering, WCU~Program, Chungbuk National University, Cheongju, South Korea 

(e-mails:~k.eshraghian@innovationlabs.com.au,~krcho@cbnu.ac.kr, omid@hbt.cbnu.ac.kr, and skkang@hbt.cbnu.ac.kr). 

O.~Kavehei is also with the School of Electrical and Electronic Engineering, University of Adelaide, SA 5005, Australia (e-mail: omid@eleceng.adelaide.edu.au).}
\thanks{D.~Abbott is with the School of Electrical and Electronic Engineering, University of Adelaide, SA 5005, Australia (e-mail:~dabbott@eleceng.adelaide.edu.au).}
\thanks{S.~M.~Kang is with the School of Engineering, University of California, Merced, CA 95343 USA (e-mail: smk123@ucmerced.edu).}
}

%
%

\markboth{IEEE TRANSACTIONS ON VERY LARGE SCALE INTEGRATION (VLSI) SYSTEMS,~Vol.~X, No.~X, ---~201X}%
{Eshraghian \MakeLowercase{\textit{et al.}}: Memristor MOS Content Addressable Memory (MCAM): Hybrid Architecture for Future High Performance Search Engines}
%



\maketitle

\begin{abstract}

Large-capacity Content Addressable Memory (CAM) is a key element in a wide variety of applications. The inevitable complexities of scaling MOS transistors introduce a major challenge in the realization of such systems. Convergence of disparate technologies, which are compatible with CMOS processing, may allow extension of Moore's Law for a few more years. This paper provides a new approach towards the design and modeling of Memristor (Memory resistor) based Content Addressable Memory (MCAM) using a combination of memristor MOS devices to form the core of a memory/compare logic cell that forms the building block of the CAM architecture. The non-volatile  characteristic and the nanoscale geometry together with compatibility of the memristor with CMOS processing technology increases the packing density, provides for new approaches towards power management through disabling CAM blocks without loss of stored data, reduces power dissipation, and has scope for speed improvement as the technology matures.  

\end{abstract}


%
\IEEEpeerreviewmaketitle

\section{Introduction}
\label{sec:Introduction}

\IEEEPARstart{T}{he} quest for a new hardware paradigm that will attain processing speeds in the order of an exaflop ($10^{18}$ floating point operations per second) and further into the zetaflop regime ($10^{21}$ flops) is a major challenge for both circuit designers and system architects. The evolutionary progress of networks such as the Internet also brings about the need for realization of new components and related circuits that are compatible with CMOS process technology as CMOS scaling begins to slow down~\cite{Bourianoff2007}. As Moore's Law becomes more difficult to fulfill, integration of significantly different technologies such as spintronics~\cite{Bourianoff2007}, carbon nano tube field effect transistors (CNFET)~\cite{Akinwande2008}, optical nanocircuits based on metamaterials~\cite{Engheta2007}, and more recently the memristor~\cite{Strukov2008}, are gaining more focus thus creating new possibilities towards realization of innovative circuits and systems within  the {\it System on System} (SoS) domain.

In this paper we explore conceptualization, design, and modeling of the memory/compare cell as part of a Memristor based Content Addressable Memory (MCAM) architecture using a combination of memristor and n-type MOS devices. A typical Content Addressable Memory (CAM) cell forms a SRAM cell that has 2 n-type and 2 p-type MOS transistors, which requires both $V_{\rm DD}$ and GND connections as well as well-plugs within each cell. Construction of a SRAM cell that exploits memristor technology, which has a non-volatile memory (NVM) behavior and can be fabricated as an extension to a CMOS process technology with nanoscale geometry, addresses the main thread of current CAM research towards reduction of power consumption.

The design of the CAM cell is based on the $4^{\rm th}$ passive circuit element, the Memristor (M) predicted by Chua in 1971~\cite{Chua1971} and generalized by Kang~\cite{Kang1975, Chua1976}. Chua postulated that a new circuit element defined by the single-valued relationship $d\phi=Mdq$ must exist, whereby current moving through the memristor is proportional to the flux of the magnetic field that flows through the material. In another words, the magnetic flux between the terminals is a function of the amount of charge, $q$, that has passed through the device. This follows from Lenz's law whereby the single-valued relationship $d\phi=Mdq$ has the equivalence $v=M(q)i$, where $v$ and $i$ are memristor voltage and current, respectively.

The memristor behaves as a switch, much like a transistor. However, unlike the transistor, it is a 2-terminal rather than a 3-terminal device and does not require power to retain either of its two states. Note that a memristor changes its resistance between two values and this is achieved via the movement of mobile ionic charge within an oxide layer, furthermore, these resistive states are non-volatile. This behavior is an important property that influences the architecture of CAM systems, where the power supply of CAM blocks can be disabled without loss of stored data. Therefore, memristor-based CAM cells have the potential for significant saving in power dissipation.

This paper has the following structure: Section~\ref{sec:Characterization-and-Modeling-Behavior-of-Memristor} is an introductory section and reviews the properties of the memristor and then explores various options available in the modeling of this device. In Section~\ref{sec:Memristor-MOS-Memory}, circuit options for realization of MCAM is investigated whereby the two disparate technologies converge to create a new CMOS-based design platform. Section~\ref{sec:simresult} provides simulation results of a basic MCAM cell to be implemented as part of a future search engine. The details of our proposed layout and preliminary CMOS overlay fabrication approach are also presented in Section~\ref{sec:layfab}. The concluding comments are provided in Section~\ref{sec:Conclusions}.

\section{Characterization and Modeling Behavior of Memristor}
\label{sec:Characterization-and-Modeling-Behavior-of-Memristor}

\begin{figure}[thpb]
\centering
\begin{tabular}{cc}
\epsfig{file=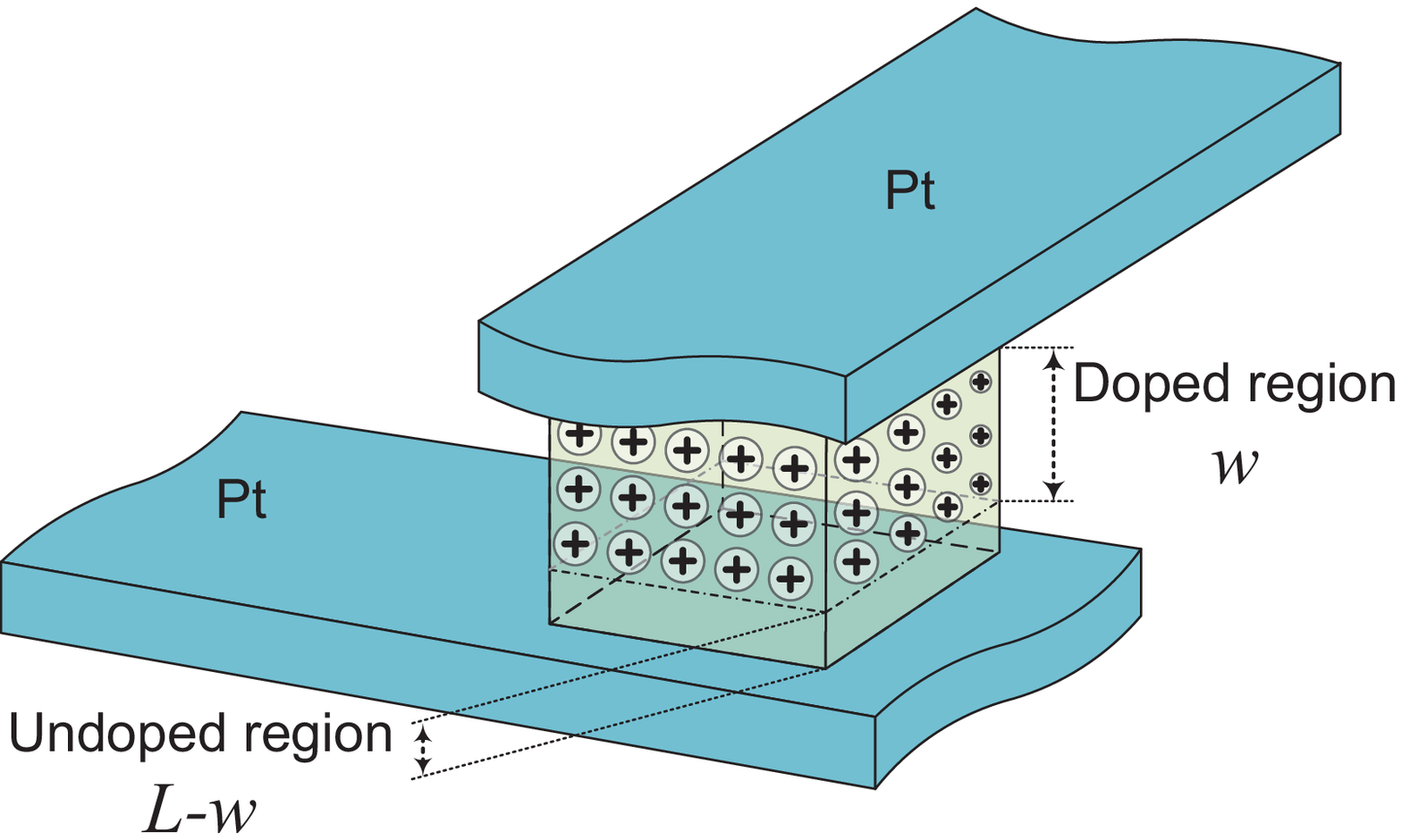,width=0.45\linewidth,clip=} & 
\epsfig{file=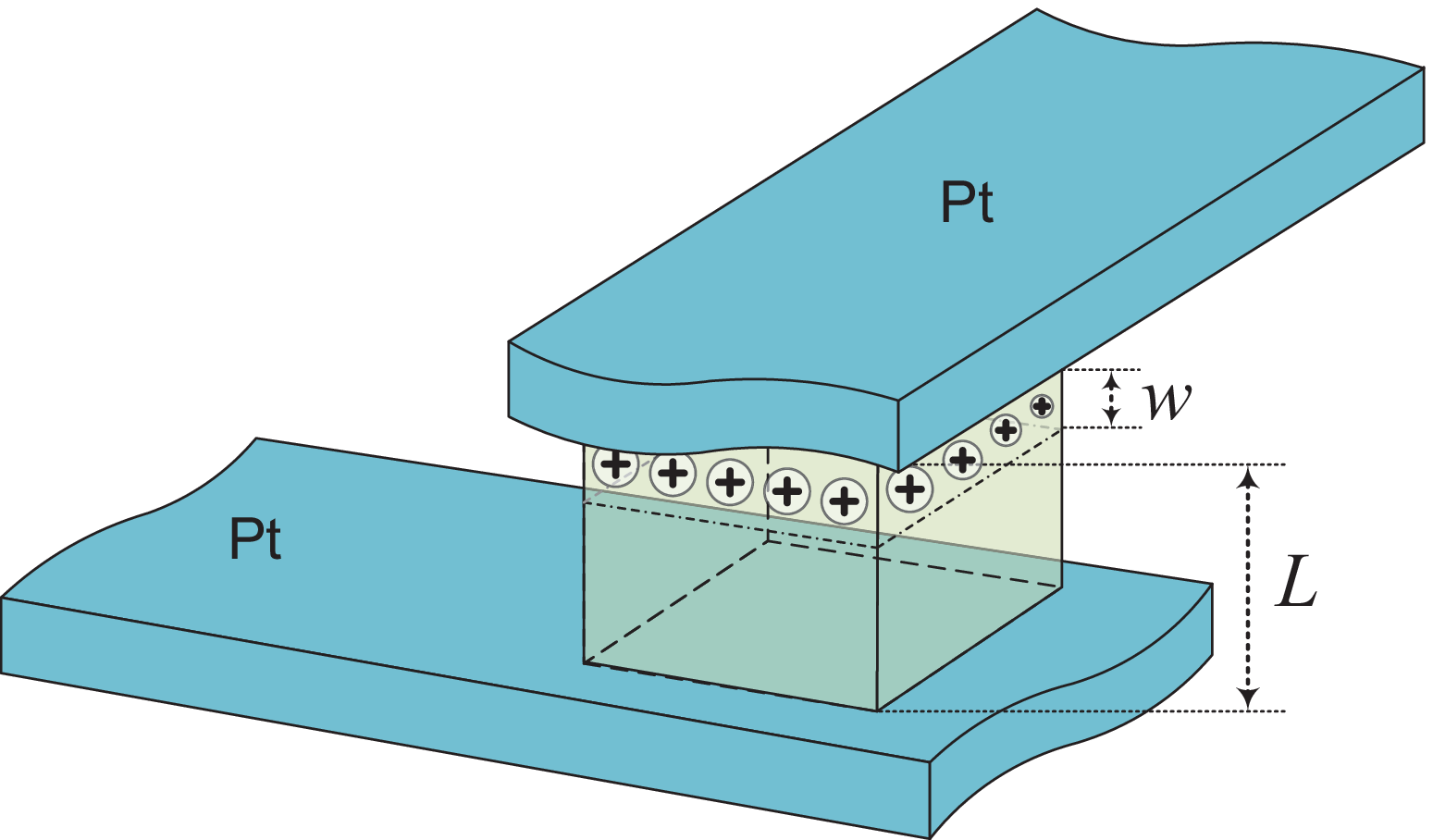,width=0.45\linewidth,clip=} 
\\ \footnotesize{(a) ``ON'' state} & \footnotesize{(b) ``OFF'' state}
\end{tabular}

  \caption{Memristor switching behavior. (a) ``ON'' state, low resistance, (b) ``OFF'' state, high resistance. The key feature of memristor is it can remember the resistance once the voltage is disconnected. In (a) ``doped'' and ``undoped'' regions are related to $R_{\rm ON}$ and $R_{\rm OFF}$, respectively. The dopant consists of mobile charges. In (b), $L$ and $w$ are the thin-film thickness and doped region thickness, respectively.}
  \label{fig:Mem-switch-behav}
\end{figure}

\citet{Strukov2008} presented a physical model whereby the memristor is characterized by an equivalent time-dependent resistor whose value at a time $t$ is linearly proportional to the quantity of charge $q$ that has passed through it. They realized a proof-of-concept memristor, which consists of a thin nano layer ($2$~nm) of TiO$_2$ and a second oxygen deficient nano layer of TiO$_{2-x}$ ($8$~nm) sandwiched between two Pt nanowires ($\sim$ 50~nm), shown in Fig.~\ref{fig:Mem-switch-behav}~\cite{Strukov2008}. Oxygen (O$^{2-}$) vacancies are +2 mobile carriers and are positively charged. A change in distribution of O$^{2-}$ within the TiO$_2$ nano layer changes the resistance. By applying a positive voltage, to the top platinum nanowire, oxygen vacancies drift from the TiO$_{2-x}$ layer to the TiO$_2$ undoped layer, thus changing the boundary between the TiO$_{2-x}$ and TiO$_2$ layers. As a consequence, the overall resistance of the layer is reduced corresponding to an ``ON'' state. When enough charge passes through the memristor that ions can no longer move, the device enters a hysteresis region and keeps $q$ at an upper bound with fixed memristance, $M$ (memristor resistance). By reversing the process, the oxygen defects diffuse back into the TiO$_{2-x}$ nano layer. The resistance returns to its original state, which corresponds to an ``OFF'' state. The significant aspect to be noted here is that only ionic charges, namely oxygen vacancies (O$^{2-}$) through the cell, change memristance. The resistance change is non-volatile hence the cell acts as a memory element that remembers past history of ionic charge flow through the cell.

	\subsection{Simplified Memristor Model}
	\label{sec:sub:Simplified-Memristor-Model}
	The memristor can be modeled in terms of two resistors in series, namely the doped region and undoped region each having vertical width of $w$ and $L-w$, respectively, as shown in Fig.~\ref{fig:Mem-switch-behav}, where $L$ is the TiO$_2$ film thickness~\cite{Strukov2008}. The voltage-current relationship defined as $M(q)$, can be modeled as~\cite{Chua1971}

\begin{equation} \label{Equ:memvolt} 
v(t) = \Bigg(R_{\rm ON}\frac{w(t)}{L}+R_{\rm OFF}\Big(1-\frac{w(t)}{L}\Big)\Bigg)i(t)~,
\end{equation}
where $R_{\rm ON}$ is the resistance for completely doped memristor, while $R_{\rm OFF}$ is the resistance for the undoped region. The width of the doped region $w(t)$ is given by,

\begin{equation} \label{Equ:memstate}
\frac{dw(t)}{dt} = \mu_v \frac{R_{\rm ON}}{L}i(t)~,
\end{equation}
where $\mu_v$ represents the average dopant mobility $\sim 10^{-10}~{\rm cm}^2/{\rm s}/{\rm V}$. Taking a normalized variable, $x(t)=w(t)/L$, instead of $w(t)$ assists in tracking memristance, $M(q)=d\phi/dq$, or memductance, $W(\phi)=dq /d\phi$. The new normalized relation is

\begin{equation} \label{Equ:memXstate}
\frac{dx(t)}{dt} = \mu_v \frac{R_{\rm ON}}{L^2}i(t)~,
\end{equation}
where $L^2 / \mu_v$ has the dimensions of magnetic flux($\phi$). Following the calculation steps from~\citet{Kavehei2009}, a simple memristance model can be defined as

\begin{equation} \label{Equ:ourmodelM}
M(t) = R_{\rm OFF}\Bigg(\sqrt{1-\frac{2c(t)}{r}}\Bigg)~,
\end{equation}
where $c(t)=\mu_v \phi(t)/L^2$, and $r$ is a ratio of $R_{\rm OFF}/R_{\rm ON}$ and $\sqrt{1-\frac{2c(t)}{r}}$ is the {\it resistance modulation index}. Here, $x(t)$ can now be rewritten as

\begin{equation} \label{Equ:ourmodelX}
x(t) = 1-\Bigg(\sqrt{1-\frac{2\phi(t)}{r\beta}}\Bigg)~,
\end{equation}
which highlights that the $r\beta$ term (where $\beta=L^2 / \mu_v$) must be made sufficiently large to maintain $2\phi(t) / r\beta$ between the range 0 and 1. The simplified linear ionic drift model facilitates the understanding of the operational characteristics of the memristor. However, for a highly nonlinear~\cite{Yang2008} relationship between electric field and drift velocity that exists at the boundaries, the ratio cannot be maintained. Thus this function is unable to model large nonlinearities close to the boundaries of the memristor characteristics. At the boundaries, i.e. when $x$ approaches 0 or 1, there is a nonlinearity associated with the memristor behavior that is discussed in the following subsection.


	\subsection{Modelling the Nonlinear Behavior of Memristor}
	\label{sec:sub:More-Complex-Model}
	The electrical behavior of the memristor as a switch/memory element is determined by the boundary between the two regions in response to an applied voltage. To model this nonlinearity, the memristor state equation Eq.~\ref{Equ:memXstate} is augmented with a {\it window function}, $F(w,i)$~\cite{Strukov2008, Strukov2009, Biolek2009, Benderli2009}, where $w$ and $i$ are the memristor's state variable and current, respectively.

\begin{figure}[thpb]
\centering
\begin{tabular}{cc}
\epsfig{file=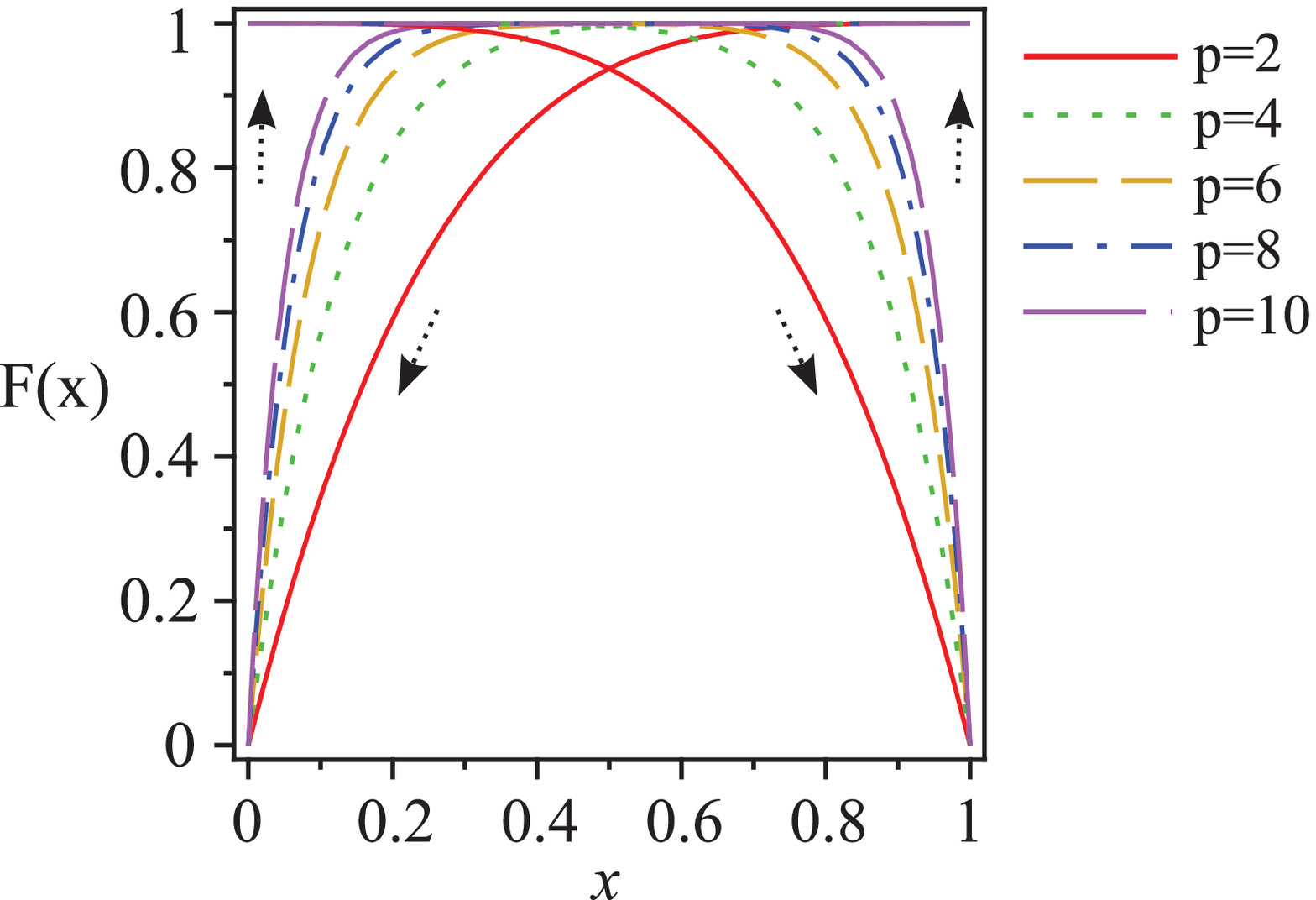,width=0.5\linewidth,clip=} &
\epsfig{file=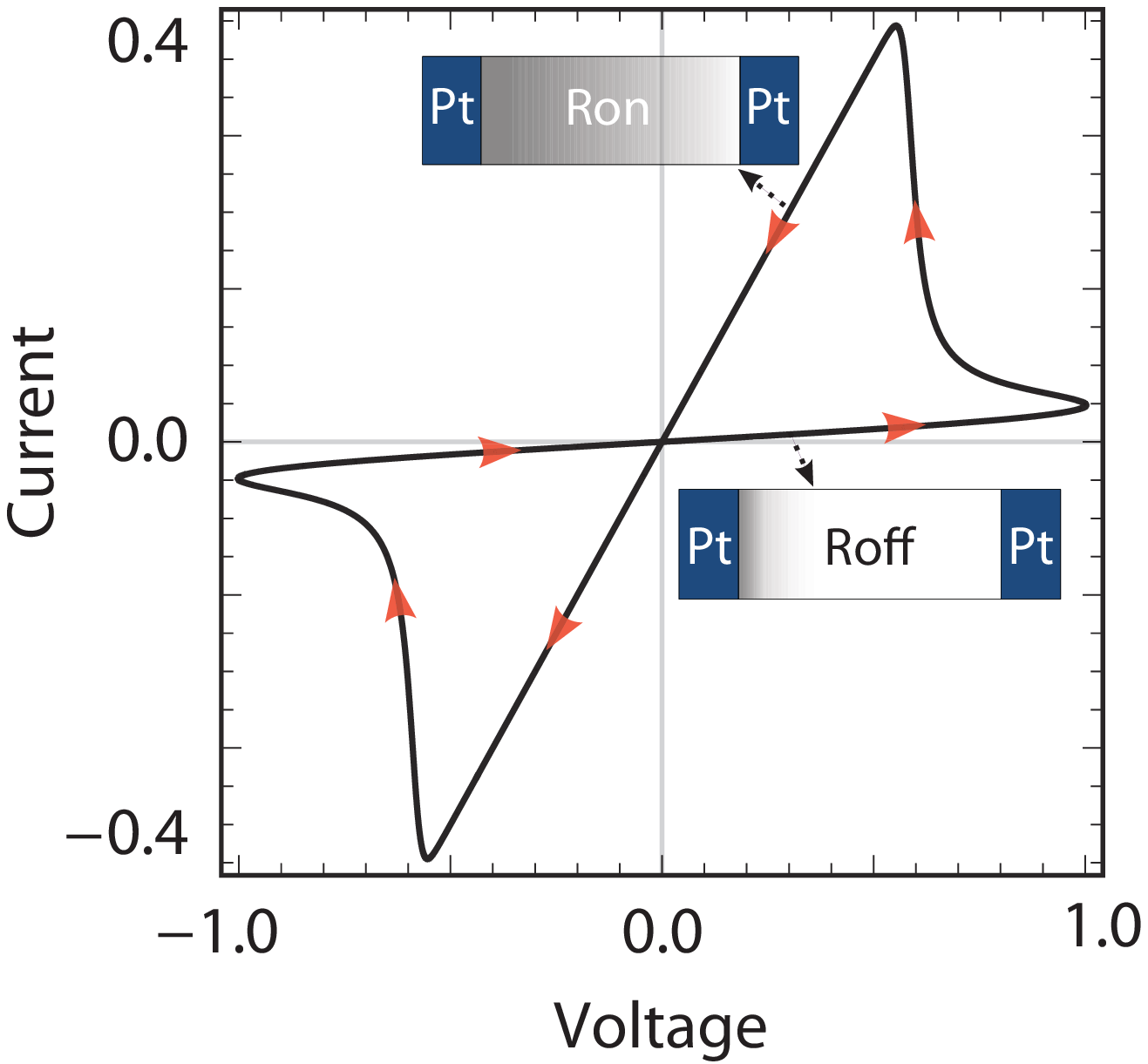,width=0.36\linewidth,clip=} \\
\footnotesize (a) Window function &
\footnotesize (b) Hysteresis characteristics
\end{tabular}
  \caption{Nonlinear behaviour of the memristor, (a) Window function: $F(x)=1-(x-{\rm sgn}(-i))^{2p}$, where ${\rm sgn}(I)$ gives the sign of the input signal $I$, (b) The hysteresis characteristics using the nonlinear drift assumption. This hysteresis shows a highly nonlinear relationship between current and voltage at the boundaries.}
  \label{fig:MemModel}
\end{figure}

 Thus, Eq.~\ref{Equ:memXstate} can be rewritten as

\begin{equation}
	\frac{dx(t)}{dt}=\frac{R_{\rm ON}}{\beta}i(t)F(x(t),p)~,
	\label{equ:nonlinstat}
\end{equation}
where $p$ is its \textit{control parameter}. The nonlinearity at the boundaries can now be controlled with parameter $p$. The influence of a window function described by Eq.~\ref{equ:nonlinstat} is illustrated in Fig.~\ref{fig:MemModel}(a) for $2\leq p\leq 10$.

\citet{Joglekar2009} proposed a modified window function to approximately address linear ionic drift and the nonlinear behaviour at the boundaries when $0<x<1$. For the window function $F(x)=1-(2x-1)^{2p}$, $p$ is a positive integer and $x=w/L$. This model considers a simple boundary condition, $F(0)=F(1)=0$, when $p\geq 4$, the state variable equation is an approximation of the linear drift assumption, $F(0<x<1)\approx 1$. This model is denoted by B-I in Table~\ref{tab:models}.

Based on this model, when a memristor is at the terminal states, no external stimulus can change its state.~\citet{Biolek2009} addressed this problem with a new window function, $F(x)=1-(x-{\rm sgn}(-i))^{2p}$, where $i$ is the memristor current, ${\rm sgn}(i)=1$ when $i\geq 0$, and ${\rm sgn}(i)=0$ when $i<0$. When current is positive, the doped region length, $w$, is expanding. This model is denoted by B-II in Table~\ref{tab:models} and is adopted for the simulations that follow. 

The hysteresis characteristic using the nonlinear drift assumption is illustrated in Fig.~\ref{fig:MemModel}(b). This hysteresis shows a highly nonlinear relationship between current and voltage at the boundaries as is derived using similar parameters reported by~\citet{Strukov2008}.

To conclude this section Table~\ref{tab:models} shows a brief comparison between different behavioral memristor models. It is also important to emphasis that the modeling approach in this paper is based on the behavioral characteristics of the solid-state thin film memristor device~\cite{Strukov2008}.~\citet{Shin2010} recently proposed compact macromodels for the solid-state thin film memristor device. Even though the assumption is still based on the linear drift model, their approach provides a solution for bypassing current flow at the two boundary resistances.

\begin{table*}[hbpt]
\centering
\caption{Comparison between different memristor models. For A-II, B-I, and B-II $x=\frac{w}{L}$.}
\begin{tabular}{lcccl}
\hline
\hline
\footnotesize & 
\footnotesize & 
\footnotesize Window Function & 
\footnotesize Boundaries & 
\footnotesize  
\\ 
\footnotesize Model & 
\footnotesize Ref & 
\footnotesize $F(\cdot)$ & 
\footnotesize $(x\rightarrow 0$, $x\rightarrow 1)$ & 
\footnotesize Problem(s) 
\\
\hline
\hline
\footnotesize A-I & 
\footnotesize \cite{Strukov2008} & 
\footnotesize $w(1-w)/L^2$ & 
\footnotesize ($0$,$\sim 0$) & 
\footnotesize Linear approximation, $x\in[0,1]$
\\ 
\footnotesize & 
\footnotesize & 
\footnotesize & 
\footnotesize & 
\footnotesize Stuck at the terminal states 
\\ 
\footnotesize & 
\footnotesize & 
\footnotesize & 
\footnotesize & 
\footnotesize $F(w\rightarrow L)\neq 0$
\\ 
\footnotesize A-II & 
\footnotesize \cite{Benderli2009} & 
\footnotesize $x(1-x)$ & 
\footnotesize ($0$, $0$) & 
\footnotesize Linear approximation, $x\in[0,1]$ 
\\ 
\footnotesize & 
\footnotesize & 
\footnotesize & 
\footnotesize & 
\footnotesize Stuck at the terminal states 
\\ 
\footnotesize B-I & 
\footnotesize \cite{Joglekar2009} & 
\footnotesize $1-(2x-1)^{2p}$ & 
\footnotesize ($0$, $0$) & 
\footnotesize Stuck at the terminal states 
\\ 
\footnotesize B-II$^*$ & 
\footnotesize \cite{Biolek2009} & 
\footnotesize $1-(x-{\rm sgn}(-i))^{2p}$ & 
\footnotesize ($0$, $0$) & 
\footnotesize Discontinuity at the boundaries 
\\ 
\hline
\hline
\end{tabular}
\\
\vspace{1 mm}
$^*$ This model is adopted for the simulations.
\label{tab:models}
\end{table*}


	\subsection{Emerging Memory Devices and Technologies}
\label{sec:sub:emrging}

Memory processing has been considered as the pace-setter for scaling a technology.   A number of performance parameters including capacity (that relate to area utilization), cost, speed (both access time and bandwidth), retention time, and persistence, read/write endurance, active power dissipation, standby power, robustness such as reliability and temperature related issues characterize memories.  Recent and emerging technologies such as Phase-Change Random Access Memory (PCRAM), Magnetic RAM (MRAM), Ferroelectric RAM (FeRAM), Resistive RAM (RRAM), and Memristor, have shown promise and some are already being considered for implementation into emerging products. Table~\ref{tab:itrs} summarizes a range of performance parameters and salient features of each of the technologies that characterize memories~\cite{ITRS2009, Freitas2008}. A projected plan for 2020 for memories highlight a capacity greater than $1$~TB, read/write access times of less than $100$~ns and endurance in the order of $10^{12}$ or more write cycles.   

\begin{table*}[htbp]
  \centering
  \caption{Traditional and emerging memory technologies}
    \begin{tabular}{l|c|c|c|c|c|c|c|c}
    \hline
	 \hline
    \multicolumn{ 1}{l}{} & \multicolumn{ 4}{c}{Traditional Technologies} & \multicolumn{ 4}{|c}{Emerging Technologies} \\
    \multicolumn{ 1}{l}{} & \multicolumn{ 2}{c}{} & \multicolumn{ 2}{|c|}{Improved Flash} & \multicolumn{ 4}{|c}{}       \\
    \multicolumn{ 1}{l|}{} & DRAM  & SRAM  & NOR   & NAND  & FeRAM & MRAM  & PCRAM   & Memristor \\
    \hline
    Knowledge level & \multicolumn{ 2}{|c|}{mature} & \multicolumn{ 2}{|c|}{advanced} & \multicolumn{ 2}{|c|}{product} & advanced & early stage \\
    \hline
    Cell Elements & 1T1C  & 6T    & \multicolumn{ 2}{|c|}{1T} & 1T1C  & 1T1R  & 1T1R  & 1M \\
    \hline
    Half pitch ($F$) (nm) & $50$    & $65$    & $90$    & $90$    & $180$   & $130$   & $65$    & $3$-$10$ \\
    \hline
    Smallest cell area ($F^2$) & $6$     & $140$   & $10$    & $5$     & $22$    & $45$    & $16$    & $4$ \\
    \hline
    Read time (ns) & $< 1$   & $< 0.3$ & $< 10$  & $< 50$  & $< 45$  & $< 20$  & $< 60$  & $< 50$ \\
    \hline
    Write/Erase time (ns) & $< 0.5$ & $< 0.3$ & $10^5$ & $10^6$ & $10$    & $20$    & $60$    & $< 250$ \\
    \hline
    Retention time (years) & seconds & N/A   & $> 10$ & $> 10$ & $> 10$ & $> 10$ & $> 10$ & $> 10$ \\
    \hline
    Write op. voltage (V) & $2.5$   & $1$     & $12$    & $15$    & $0.9$-$3.3$ & $1.5$   & $3$     & $< 3$ \\
    \hline
    Read op. voltage (V) & $1.8$   & $1$     & $2$  & $2$  & $0.9$-$3.3$ & $1.5$   & $3$     & $< 3$ \\
    \hline
    Write endurance & $10^{16}$ & $10^{16}$ & $10^5$ & $10^5$ & $10^{14}$ & $10^{16}$ & $10^9$ & $10^{15}$ \\
    \hline
    Write energy (fJ/bit) & $5$ & $0.7$ & $10$ & $10$ & $30$ & $1.5\times 10^{5}$ & $6\times 10^{3}$ & $< 50$ \\
    \hline
    Density (Gbit/cm$^2$) & $6.67$  & $0.17$  & $1.23$  & $2.47$  & $0.14$  & $0.13$  & $1.48$  & $250$ \\
	\hline
    Voltage scaling & \multicolumn{ 5}{|c|}{fairly scalable} & no & poor  & promising \\
   \hline
    Highly scalable  & \multicolumn{ 4}{|c|}{major technological barriers} & \multicolumn{ 2}{|c|}{poor} & promising  & promising \\
   \hline
	\hline
    \end{tabular}
  \label{tab:itrs}
\end{table*}

Flash memories suffer from both a slow write/erase times and low endurance cycles. FeRAMs and MRAMs are poorly scalable. MRAMs and PCRAMs require large programming currents during write cycle, hence an increase in dissipation per bit. Furthermore, voltage scaling becomes more difficult. Memristors, however, have demonstrated promising results in terms of the write operation voltage scaling~\cite{Strukov2009, Kuekes2005}.

Memristor crossbar-based architecture is highly scalable~\cite{Strukov2009b} and shows promise for ultra-high density memories~\cite{Vontobel2009}. For example, a memristor with minimum feature sizes of $10$~nm and $3$~nm yield $250$~Gb/cm$^2$ and $2.5$ Tb/cm$^2$, respectively.   

In spite of the high density, zero standby power dissipation, and long life time that have been pointed out for the emerging memory technologies, their long write latency has a large negative source of impact on memory bandwidth, power consumption, and the general performance of a memory system.


\section{Conventional CAM and the Proposed MCAM Structures}
\label{sec:Memristor-MOS-Memory}

\begin{figure}[htbp]
  \begin{center}
	\includegraphics[angle=0,scale=0.22]{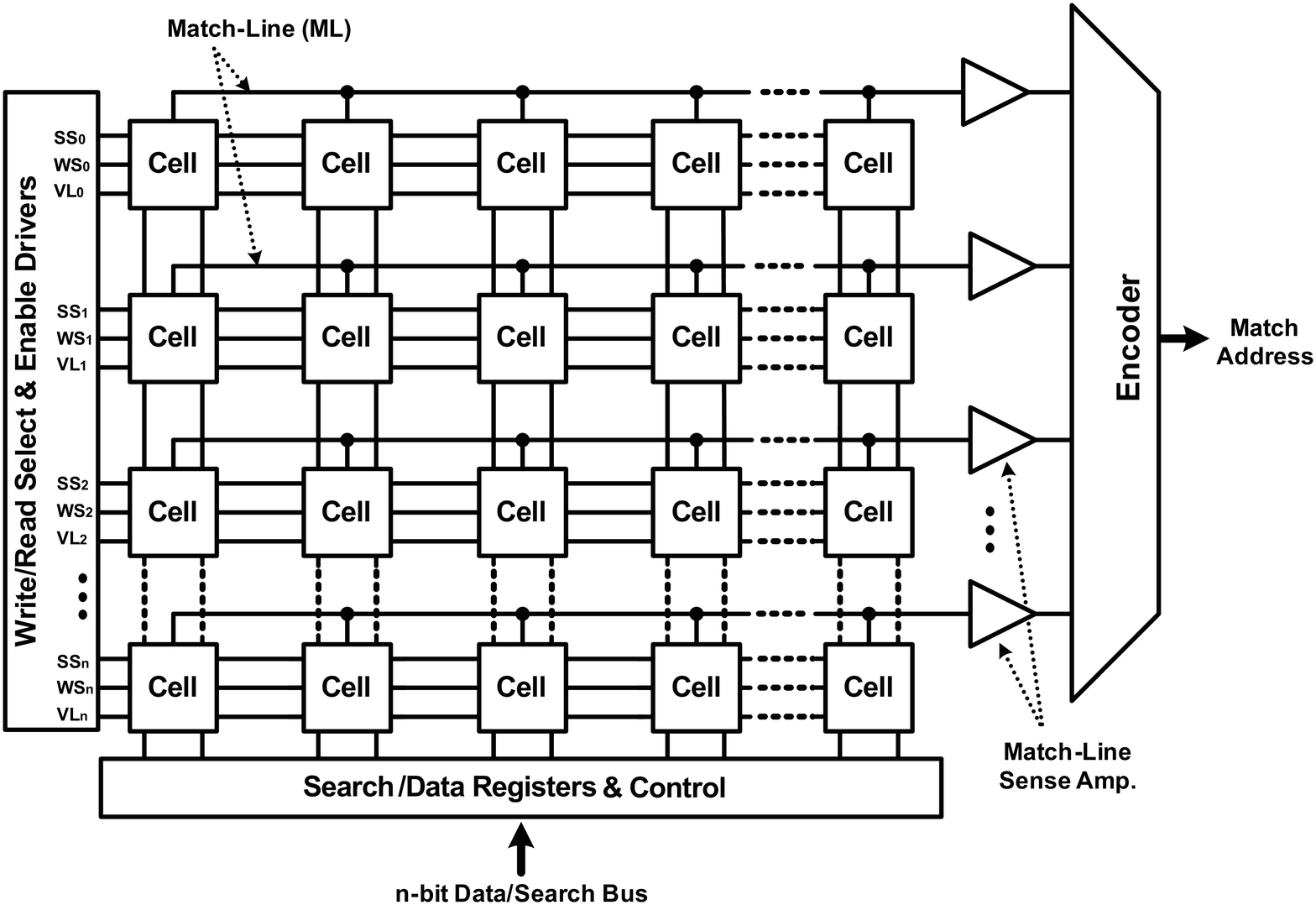}
  \end{center}

  \caption{Generic Content Addressable Memory Architecture for $n\times n$ NAND-type CAM cells. In this structure each data (D) and search (S) bits share one common bus line (D/S) to reduce the interconnection complexity. The architecture is based on the MCAM cell of Fig.~\ref{fig:CAMcells}(d) and the match lines (MLs) composed of nMOS pass transistors.}  
  \label{fig:genarch}
\end{figure}

A content addressable memory illustrated in Fig.~\ref{fig:genarch} takes a search word and returns the matching memory location. Such an approach can be considered as a mapping of the large space of input search word to that of the smaller space of output match location in a single clock cycle~\cite{Tyshchenko2008}. There are numerous applications including Translation Lookaside Buffers (TLB), image coding~\cite{Kumaki2007}, classifiers to forward Internet Protocol (IP) packets in network routers~\cite{Kim2009b}, etc. Inclusion of memristors in the architecture ensures that data is retained if the power source is removed enabling new possibilities in system design including the all important issue of power management.

\subsection{Conventional Content Addressable Memory}
\label{sec:sub:sub:convcam}

\begin{figure}[thpb]
\centering
\begin{tabular}{cc}
\epsfig{file=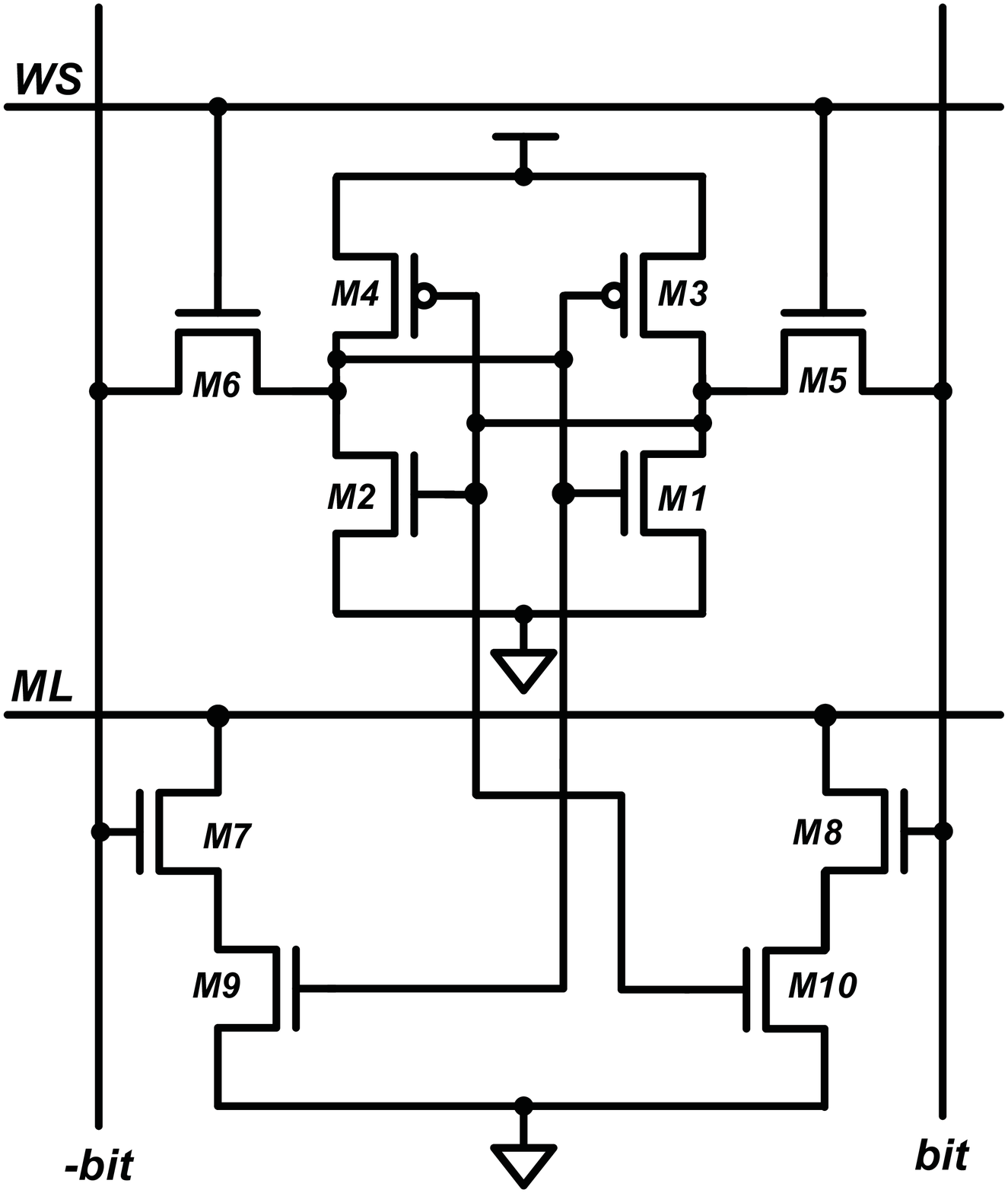,width=0.45\linewidth,clip=} \\
\footnotesize{(a) Conventional 10-T NOR-type CAM Cell} \\
\epsfig{file=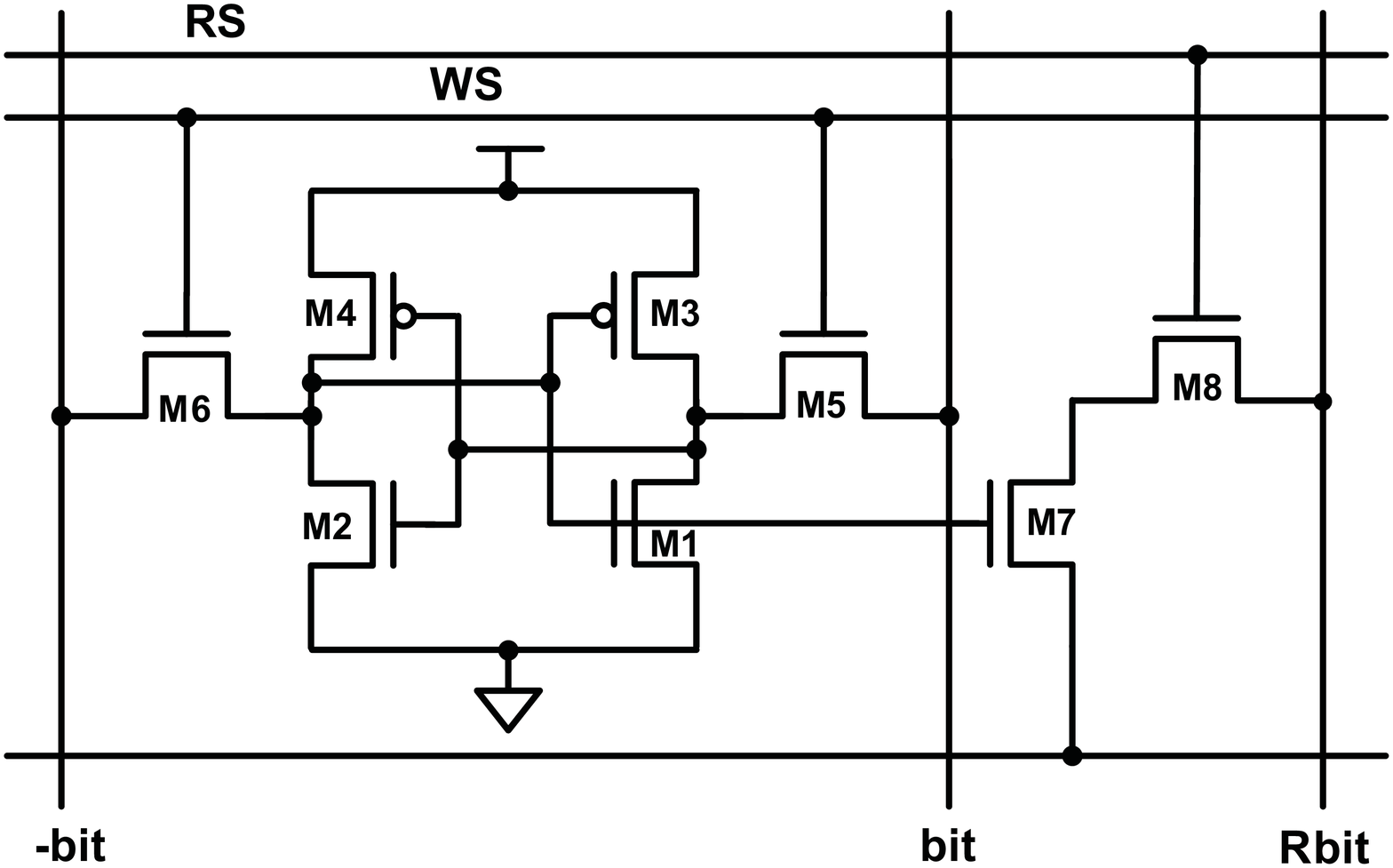,width=0.65\linewidth,clip=} \\
\footnotesize{(b) 8-T Subthreshold SRAM Cell~\cite{Verma2008}}
\end{tabular}

  \caption{Conventional CAM cell structure and the design of a SRAM cell for ultra low-power applications. In (a) a conventional 10-T NOR-type CAM circuit is demonstrated. Usually, conventional NOR- or NAND-type CAM cells have more than $9$ transistors~\cite{Pagiamtzis2006}. In (a) and (b), RS, Rbit, WS, ML, bit, and -bit lines are read select, read bit-line, word select, match line, data, and complementary data signals.}
  \label{fig:convcam}
\end{figure}

To better appreciate some of the benefits of our proposed structure we provide a brief overview of the conventional CAM cell using static random access memory (SRAM) as shown in Fig.~\ref{fig:convcam}(a). The two inverters that form the latch use four transistors including two p-type transistors that normally require more silicon area. Problems such as relatively high leakage current particularly for nanoscaled CMOS technology~\cite{Verma2008} and the need for inclusion of both $V_{\rm DD}$ and ground lines in each cell bring further challenges for CAM designers in order to increase the packing density and still maintain sensible power dissipation. Thus, to satisfy the combination of ultra dense designs, low-power (low-leakage), and high-performance, the SRAM cell is the focus of architectural design considerations. 

For instance, one of the known problems of the conventional 6-T SRAM for ultra low-power applications is its static noise margin (SNM)~\cite{Verma2008}. Fundamentally, the main technique used to design an ultra low-power memory is voltage scaling that brings CMOS operation down to the subthreshold regime.~\citet{Verma2008} demonstrated that at very low supply voltages the static noise margin for SRAM will disappear due to process variation. To address the low SNM for subthreshold supply voltage~\citet{Verma2008} proposed 8-T SRAM cell shown in Fig.~\ref{fig:convcam}(b). This means, there is a need for significant increase in silicon area to have reduced failure when the supply voltage has been scaled down.

Failure is a major issue in designing ultra dense (high capacity) memories. Therefore, a range of fault tolerance techniques are usually applied~\cite{Lu2006}. As long as the defect or failure results from the SRAM structure, a traditional approach such as replication of memory cells can be implemented. Obviously it causes a large overhead in silicon area which, exacerbates the issue of power consumption.

Some of the specific CAM cells, for example, ternary content addressable memory (TCAM) normally used for the design of high-speed lookup-intensive applications in network routers, such as packet forwarding and classification two SRAM cells, are required. Thus, the dissipation brought about as the result of leakage becomes a major design challenge in TCAMs~\cite{Mohan2009}. It should be noted that the focus in this paper is to address the design of the store/compare core cell only, leaving out details of CAM's peripherals such as read/write drivers, encoder, matchline sensing selective precharge, pipelining, matchline segmentation, current saving technique etc., that characterize a CAM architecture~\cite{Pagiamtzis2006}.

	\subsection{Generic Memristor-nMOS Circuit}
\label{sec:sub:genericcir}

\begin{figure}[thpb]
\centering
\begin{tabular}{cc}
\epsfig{file=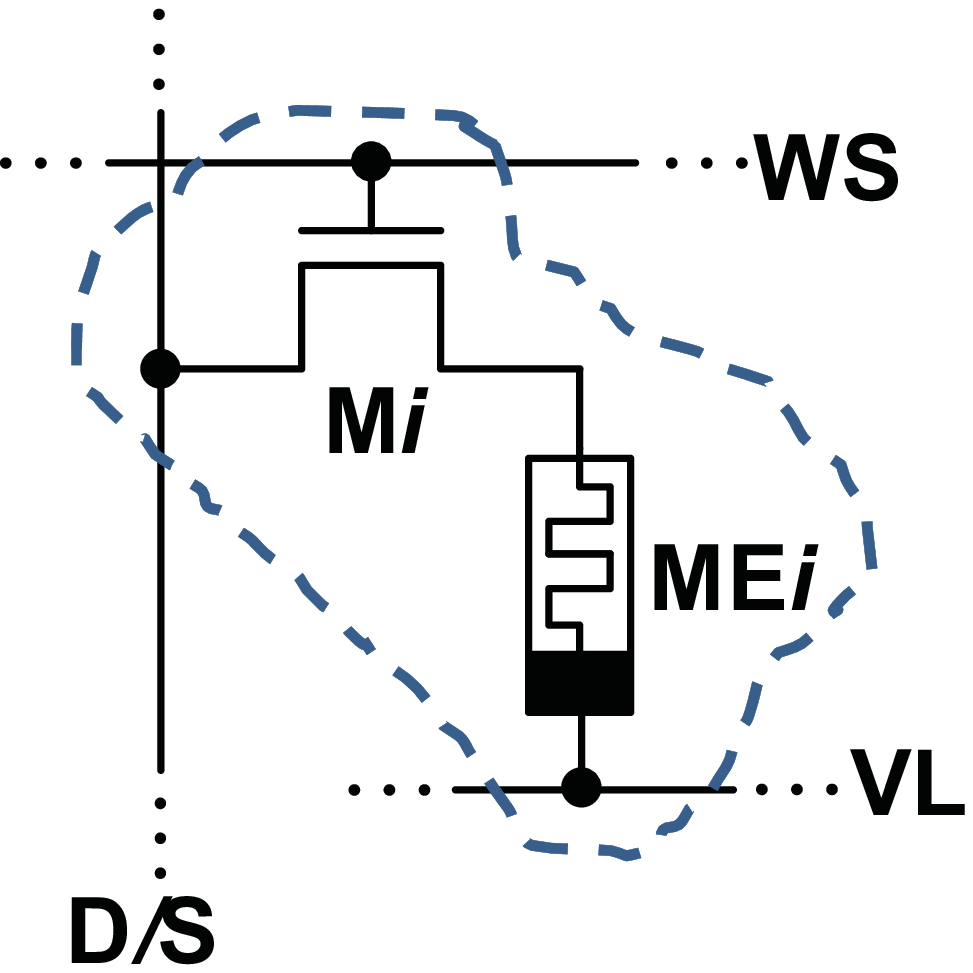,width=0.25\linewidth,clip=} &
\epsfig{file=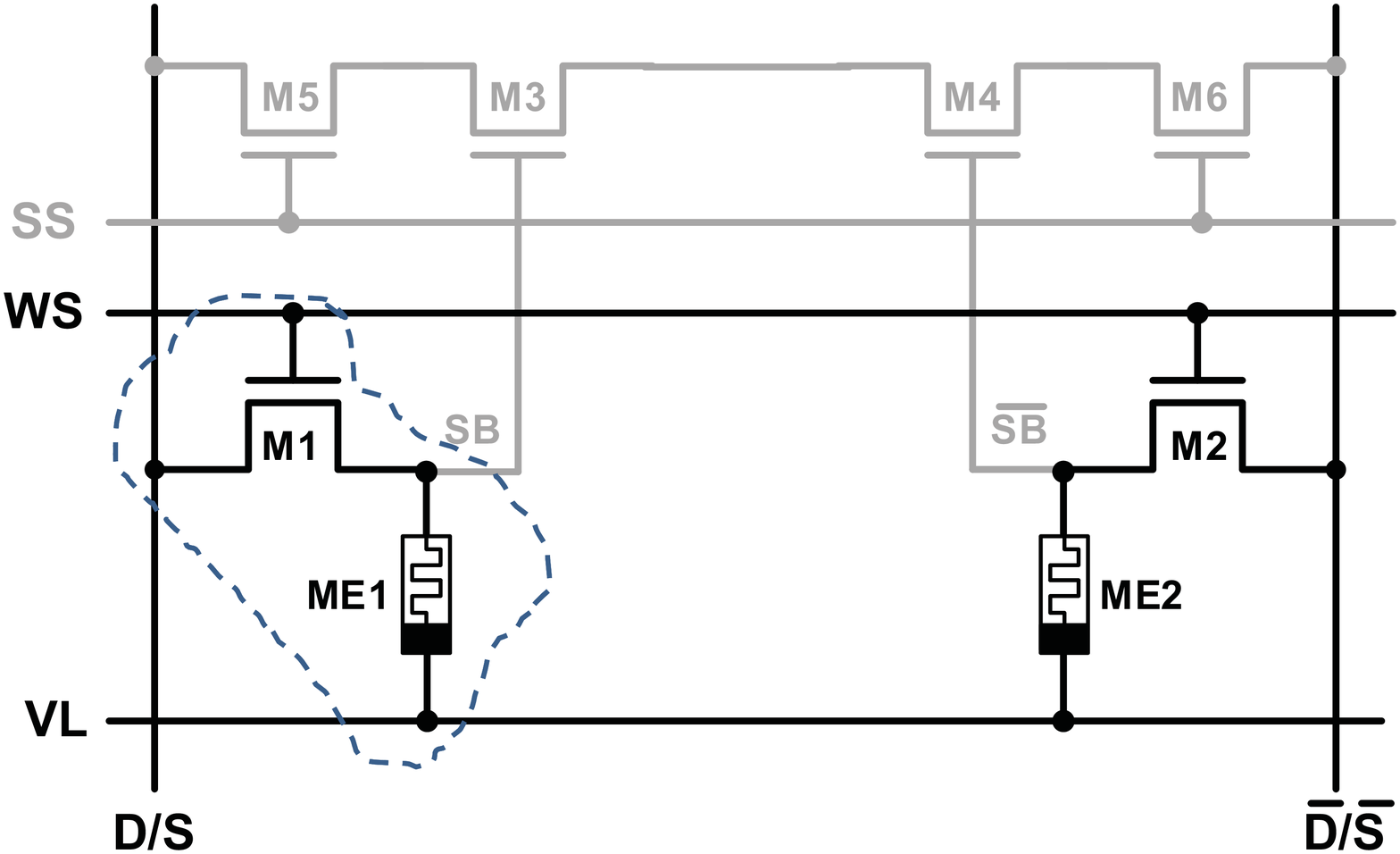,width=0.48\linewidth,clip=} \\ 
\footnotesize{(a) Structure of write mode.} & \footnotesize{(b) Basic cell.} \\
\epsfig{file=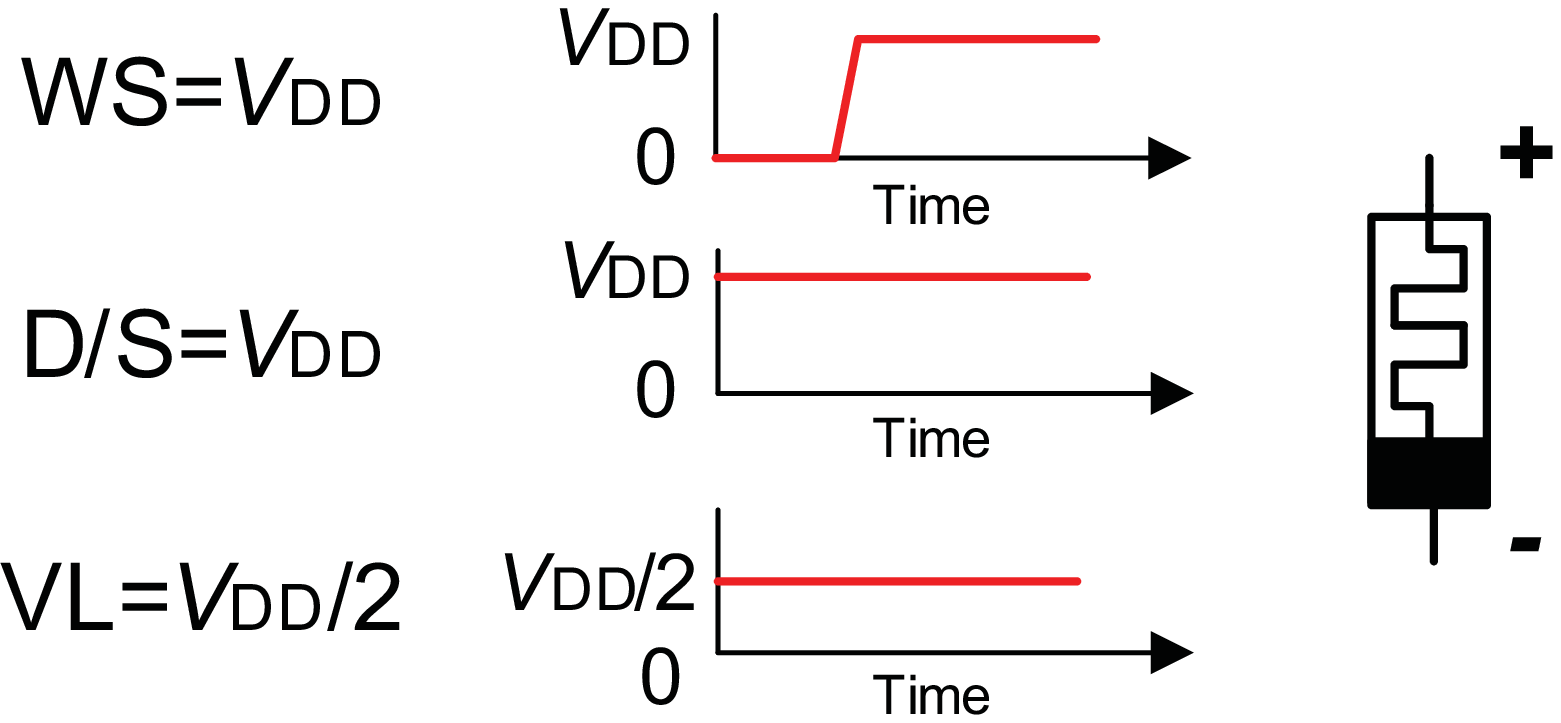,width=0.45\linewidth,clip=} &
\epsfig{file=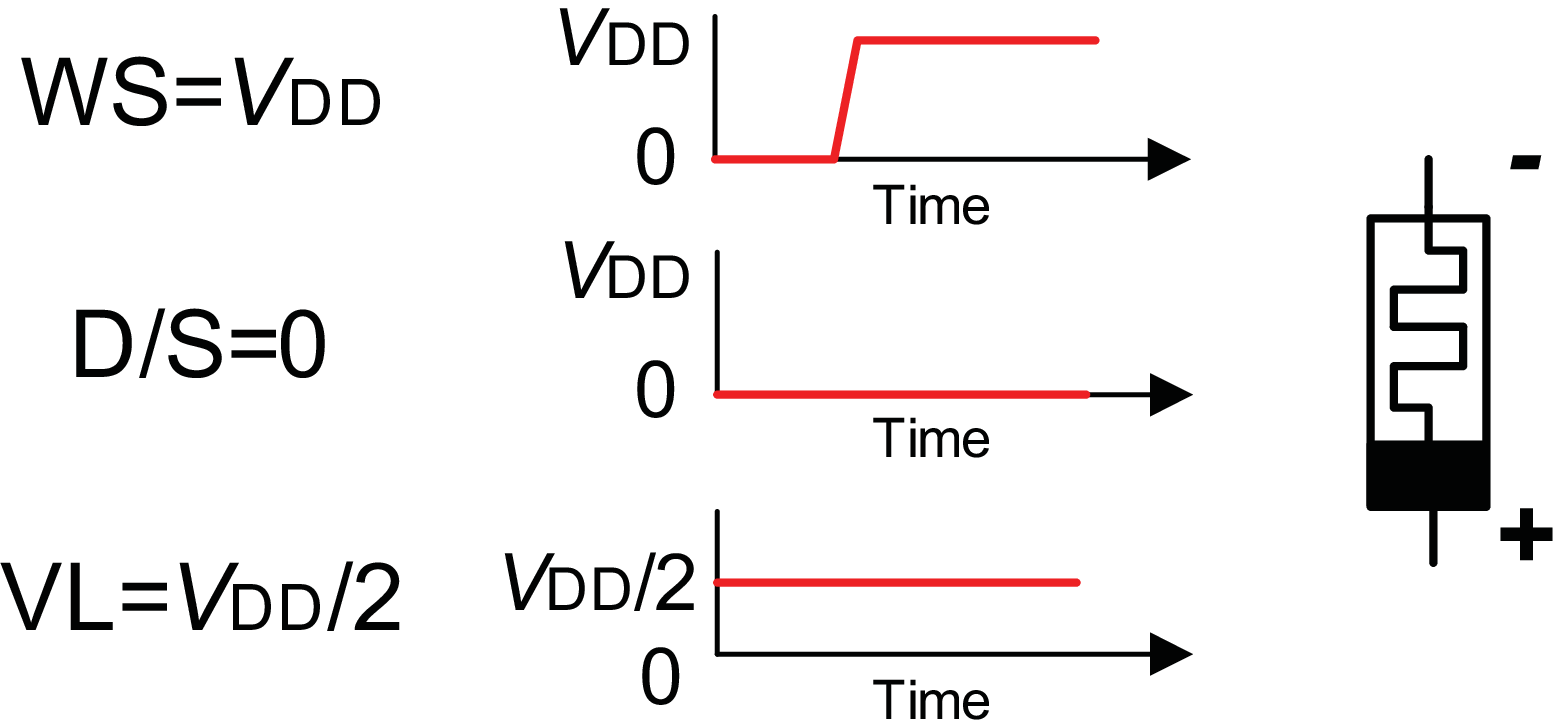,width=0.45\linewidth,clip=} \\ 
\footnotesize{(c) Program  ``Low'' resistance ``1''.} & \footnotesize{(d) Program  ``High'' resistance ``0''.}
\end{tabular}

  \caption{Basic memristor-nMOS storage cell and the timing diagram. (a) shows write mode part of the $i$-th cell in a row. (b) Basic cell circuit without the match-line transistor. (c) ``Low'' resistance, $R_{\rm ON}$, programing. Equivalent to logic ``1''. (d) ``High'' resistance, $R_{\rm OFF}$, programing. Equivalent to logic ``0''.}
  \label{fig:basicm}
\end{figure}

Fig.~\ref{fig:basicm} shows the basic structure for a memristor-nMOS storage cell. For writing a logic ``1'', the memristor receives a positive bias to maintain an ``ON'' state. This corresponds to the memristor being programmed as a logic ``1''. To program a ``0'' a reverse bias is applied to the memristor, which makes the memristor resistance high. This corresponds to logic ``0'' being programmed.


	\subsection{MCAM Cell}
\label{sec:sub:Memristor-MOS-based-Content-Addressable-Memory-Cell-Architecture}
	In this subsection, variations of MCAM cells as well as a brief architectural perspective are introduced. The details of read/write operations and their timing issues are also discussed in the next section. A CAM cell serves two basic functions: ``bit storage'' and ``bit comparison''. There are a variety of approaches in the design of basic cell such as NOR based match line, NAND based match line, etc. This part of the paper reviews the properties of conventional SRAM-based CAM and provides a possible approach for the design of content addressable memory based on the memristor.



	\subsubsection{MCAM Cell Properties}
\label{sec:sub:sub:mcam}

\begin{figure}[thpb]
\centering
\begin{tabular}{cc}
\epsfig{file=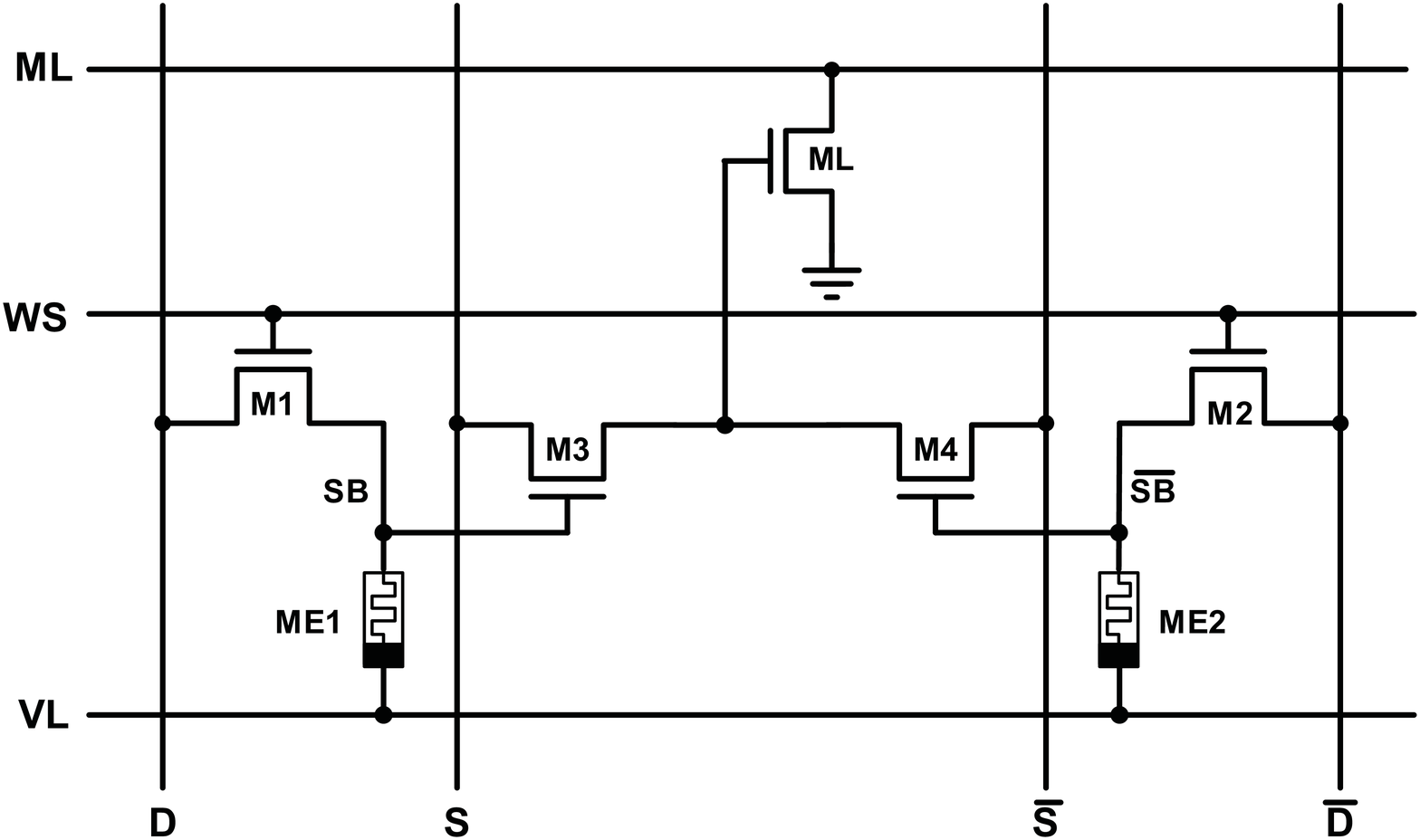,width=0.7\linewidth,clip=} \\ 
\footnotesize{(a) 5-T NOR-type} \\
\epsfig{file=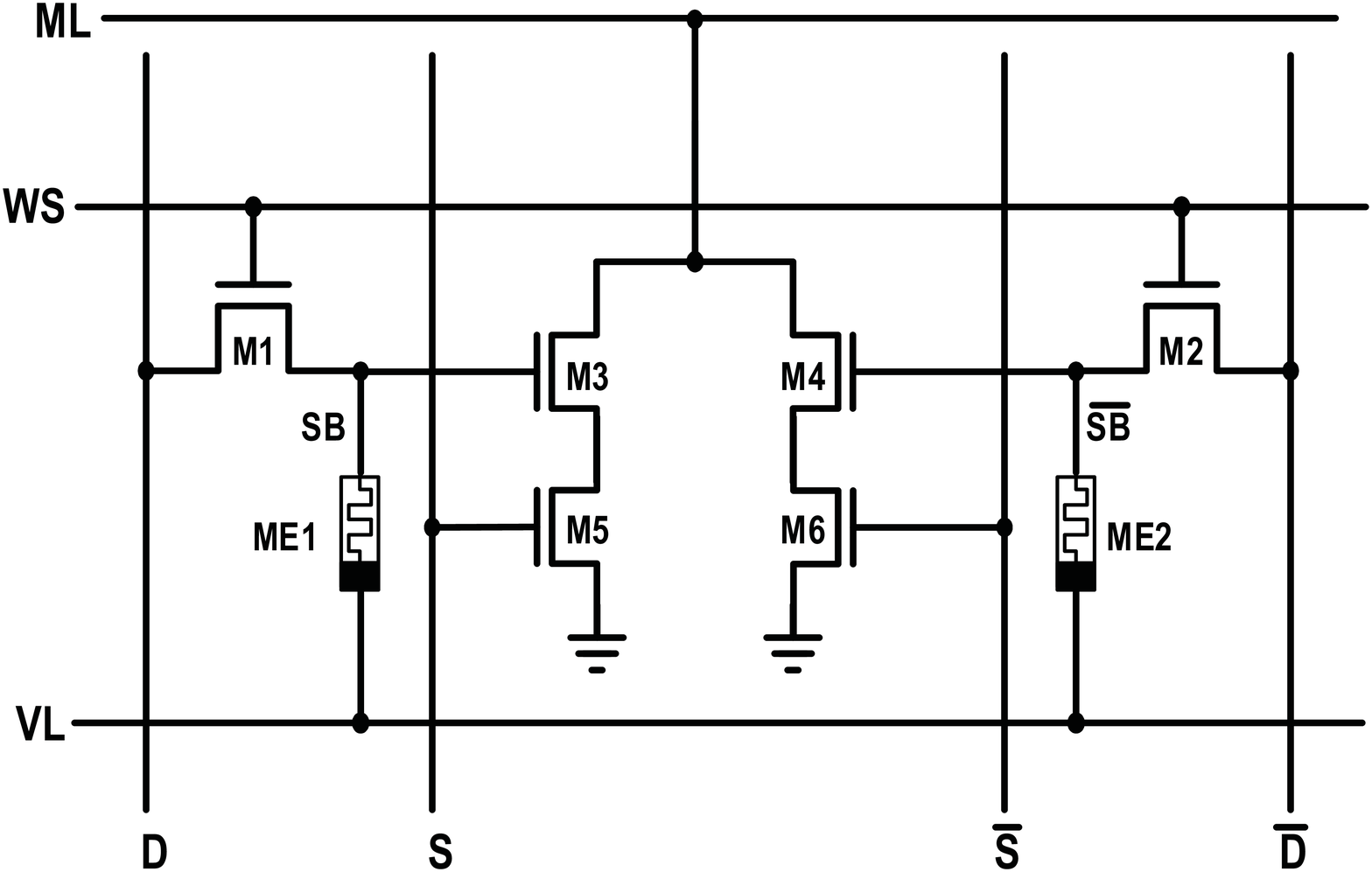,width=0.7\linewidth,clip=} \\
\footnotesize{(b) 6-T NOR-type} \\
\epsfig{file=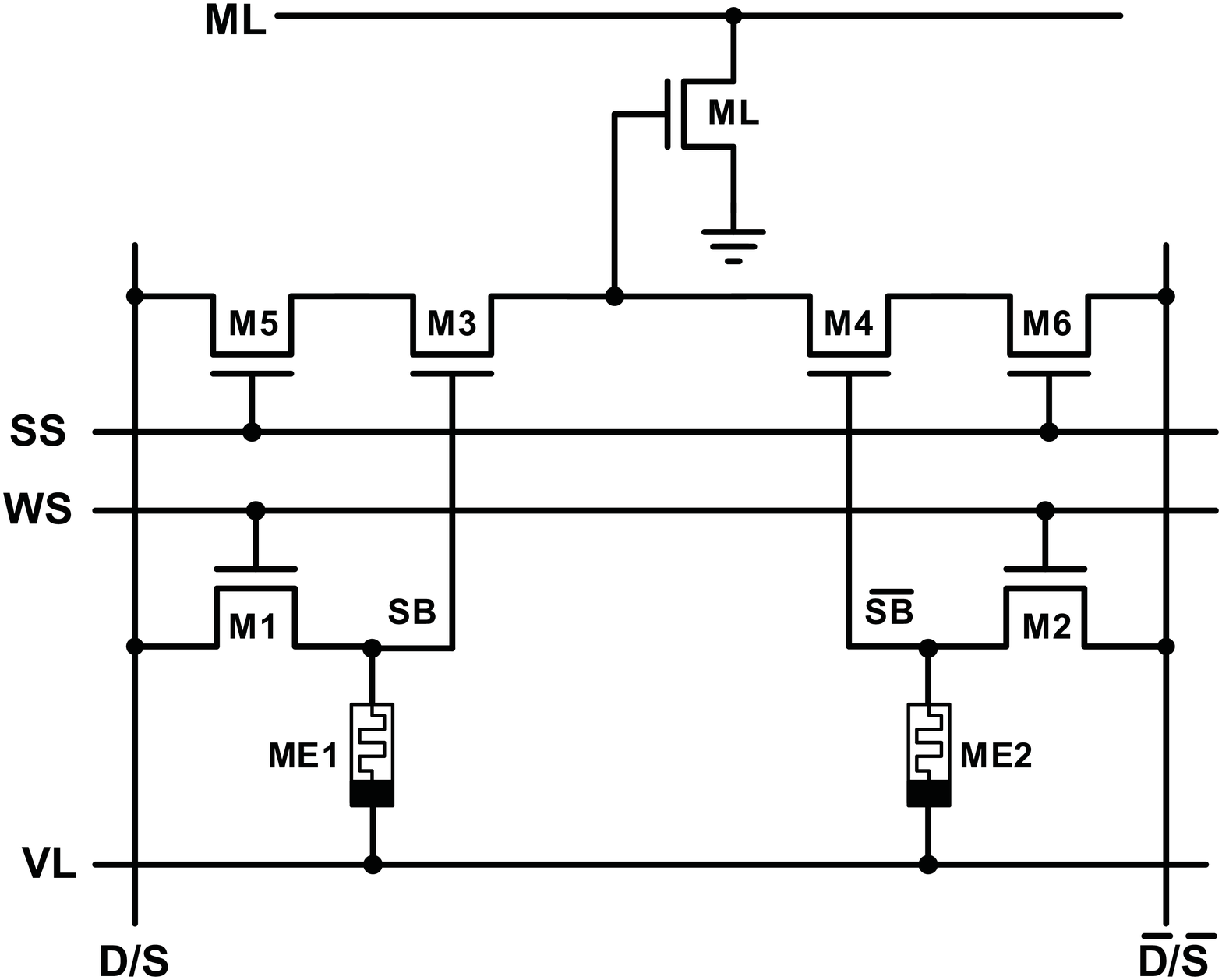,width=0.65\linewidth,clip=} \\
\footnotesize{(c) 7-T NOR-type} \\
\epsfig{file=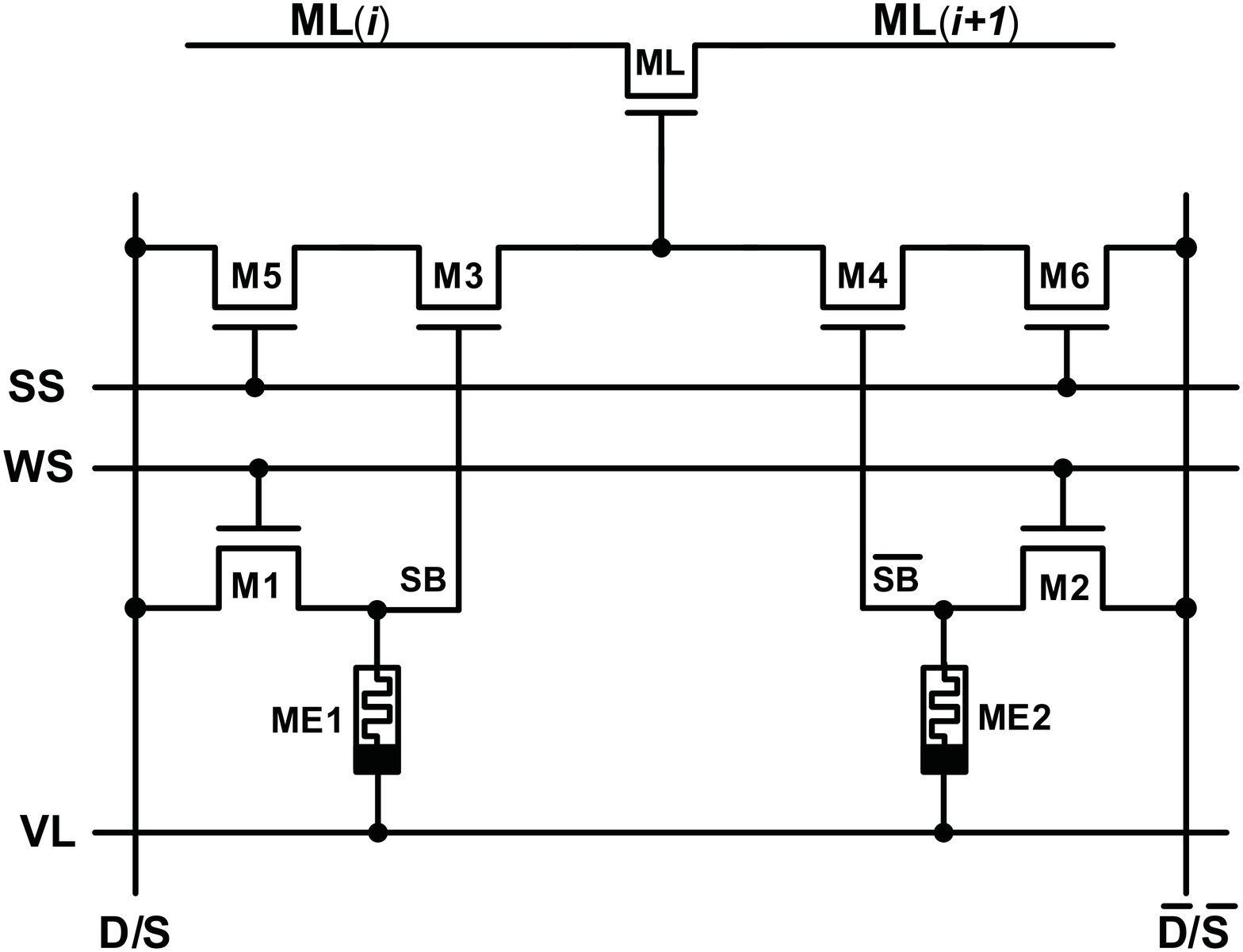,width=0.65\linewidth,clip=} \\
\footnotesize{(d) 7-T NAND-type} 
\end{tabular}

  \caption{Cell configurations of possible MCAM structures.}
  \label{fig:CAMcells}
\end{figure}

Fig.~\ref{fig:CAMcells} illustrates several variations of the MCAM core whereby bit-storage is implemented by memristors ME1 and ME2. Bit comparison is performed by either NOR or alternatively NAND based logic as part of the match-line ML$_i$ circuitry. The matching operation is equivalent to logical XORing of the search bit (SB) and stored bit (D). The match-line transistors (ML) in the NOR-type cells can be considered as part of a pull-down path of a  pre-charged NOR gate connected at the end of each individual ML$_i$ row. The NAND-type CAM functions in a similar manner forming the pull-down of a pre-charged NAND gate. Although each of the selected cells in Fig.~\ref{fig:CAMcells} have their relative merits, the approach in Fig.~\ref{fig:CAMcells}(c) where Data bits and Search bits share a common bus is selected for detailed analysis. The structure of the 7-T NAND-type, shown in  Fig.~\ref{fig:CAMcells}(d), and the NOR-type are identical except for the position of the ML transistor. In the NOR-type, ML makes a connection between shared ML and ground while in the NAND-type, the ML transistors act as a series of switches between the ML$_{i}$ and ML$_{i+1}$. 

	\section{Simulation Results Analysis and Comparison}
\label{sec:simresult}

Generally, there are the ``write'' and ``read'' operations that require consideration. In this section the ``write'' and ``read'' operations of the basic MCAM cell for 7-T NOR-type are reported. Simulations of the circuits are based on the following parameters~\cite{Witrisal2009a}: $R_{\rm ON}=100~\Omega$, $R_{\rm OFF}=100~{\rm k}\Omega$, $p=4$, $L=3~{\rm nm}$, and $\mu_v=3\times 10^{-8}~{\rm m^2/s/V}$. Both the conventional CAM and MCAM circuits have been implemented using Dongbu HiTech $0.18~\mu{\rm m}$ technology where $1.8$ Volts is the nominal operating voltage for the CAM. The MCAM cell is implemented using nMOS devices and memristors without the need for $V_{\rm DD}$ voltage source. Using the above memristor parameters, together with the behavioral model B-II of Table~\ref{tab:models}, satisfactory operation of the MCAM cell is achieved at $3.0$ Volts. We have referred to this voltage as the nominal voltage for the MCAM cell. Furthermore, the initial state of the memristors (``ON'', ``OFF'', or in between) is determined by initial resistance, $R_{\rm INIT}$. 


	\subsection{Write operation}
\label{sec:sub:write}

\begin{figure}[thpb]
\centering
\begin{tabular}{cc}
\epsfig{file=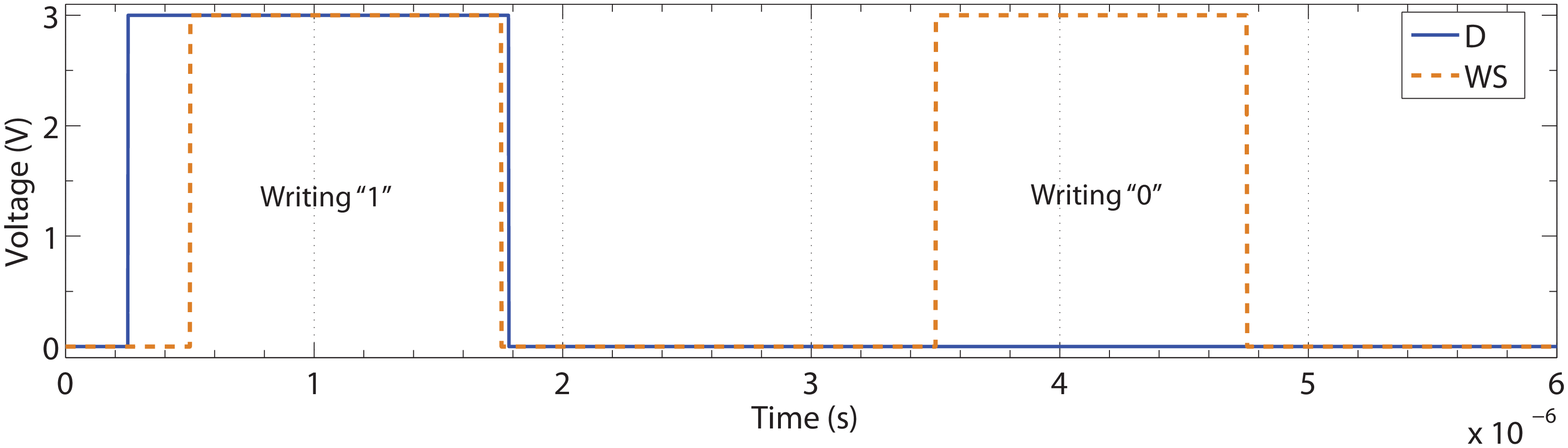,width=0.95\linewidth,clip=} \\
 \footnotesize{(a) Data (D) and Word Select (WS) signals. WS pulse width is $1.2~\mu s$.} \\
\epsfig{file=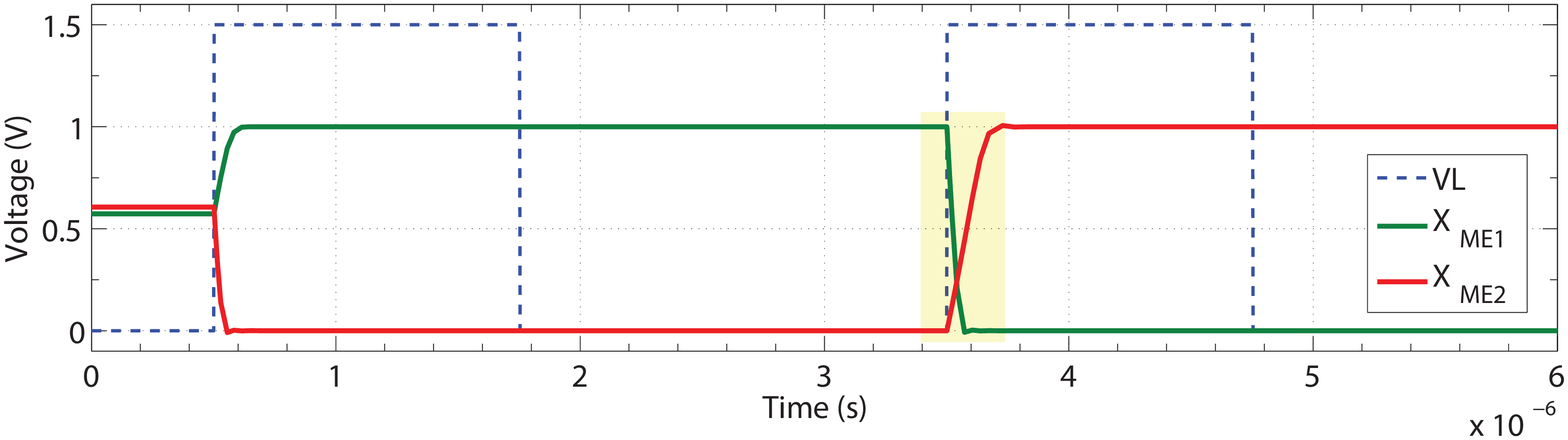,width=0.95\linewidth,clip=} \\ 
\footnotesize{(b) Write enable, VL, and memristors state, $x_{\rm ME1}$ and $x_{\rm ME2}$, signals.}
\end{tabular}

  \caption{Write operation timing diagram. The highlighted area in (b) shows the minimum time for writing, which is the maximum for both memristors, around $220$~ns. In (b) $x_{\rm ME1}$ and $x_{\rm ME2}$ are dimensionless parameters and both are varying between $0$ and $1$. The rational for showing VL and $x_{\rm ME1}$ and $x_{\rm ME2}$ together is that VL acts as a trigger for the state variables. VL$_{\rm active}=1.5$ V ($V_{\rm DD}/2$) for write operation.}
  \label{fig:writeph}
\end{figure}

At the write phase, the memristor ME1 is programmed based on the data bit on the D line. The complementary data is also stored in ME2. During the write operation, the select line is zero and an appropriate write voltage is applied on VL. The magnitude of this voltage is half of supply voltage, that corresponds to $V_{\rm DD}/2$. The pulse width is determined by the time required for the memristor to change its state from logic ``1'' ($R_{\rm ON}$) to logic ``0'' ($R_{\rm OFF}$) or vice versa. Waveforms in Fig.~\ref{fig:writeph} illustrate the write operation. In this case $R_{\rm INIT}=40~{\rm k}\Omega$ and the initial state is around $0.6$. The diagrams show two write operations, for both when D is ``1'' and when it is ``0''. By applying $V_{\rm DD}/2$ to VL line, there will be a $-V_{\rm DD}/2$ potential across the memristor ME2 and $V_{\rm DD}-V_{\rm th,M1}$ across the memristor ME1.

The highlighted area in Fig.~\ref{fig:writeph}(b) shows the difference in the write operation between ME1 and ME2. When ${\rm D}=0$ and $\overline{\rm D}=V_{\rm DD}$, there is a threshold voltage ($V_{\rm th}$) drop at the $\overline{\rm SB}$ node. Thus, the potential across the memristor would be $V_{\rm DD}/2-V_{\rm th,M2}$. At the same time, $-V_{\rm DD}/2$ is the voltage across the ME1, so the change in state in ME1 occurs faster than memristor ME2. The time for a state change is approximately $75$~ns for ME1 and $220$~ns for ME2. Therefore, $145$~ns delay is imposed because of the voltage drop across the ME2. Fig.~\ref{fig:writeph}(b) illustrates simulation results carried out using a behavioral SPICE macro-model.

	\subsection{Read operation}
\label{sec:sub:read}

\begin{figure*}[thpb]
\centering
\begin{tabular}{cc}
\epsfig{file=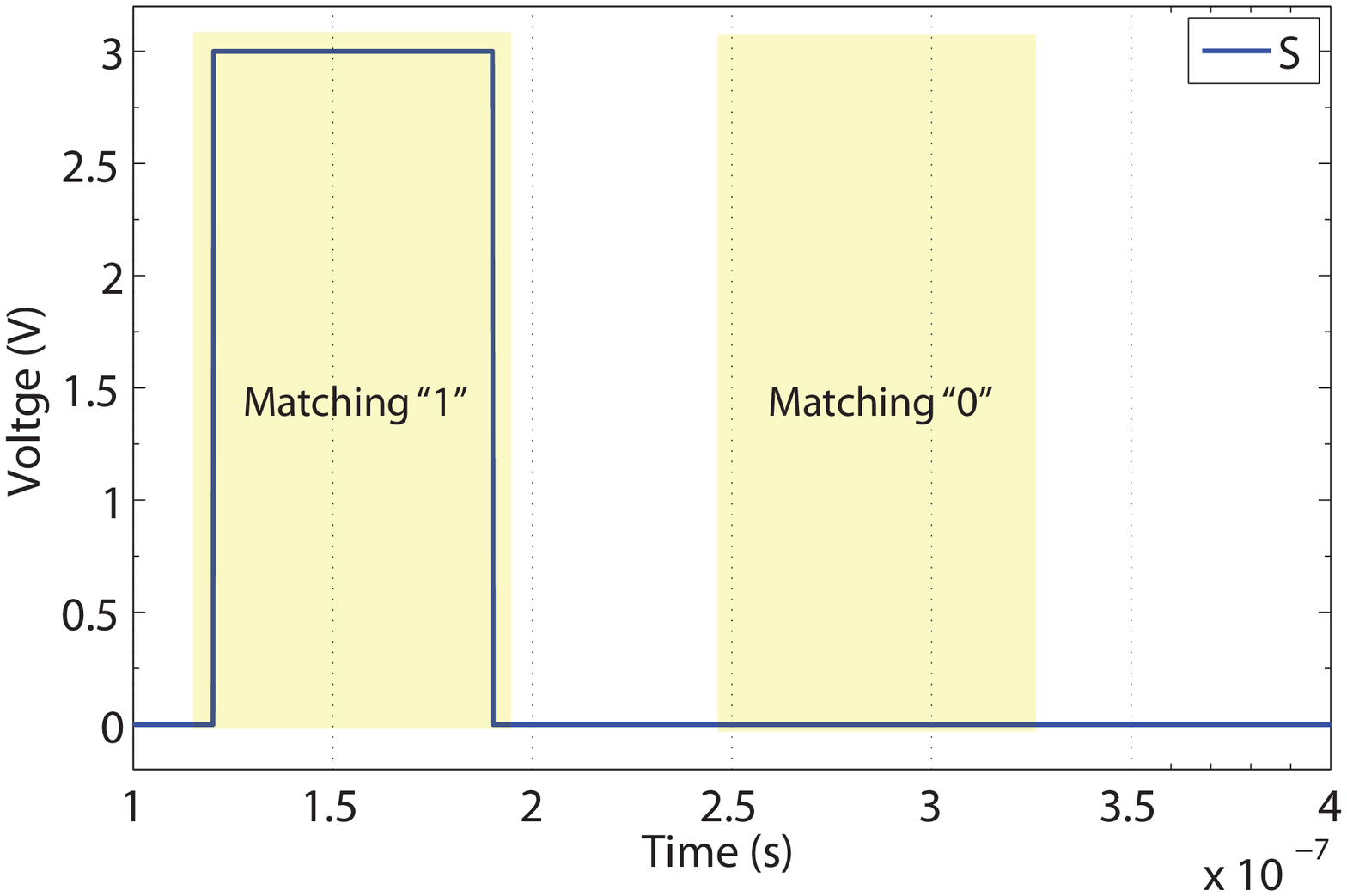,width=0.4\linewidth,clip=} & 
\epsfig{file=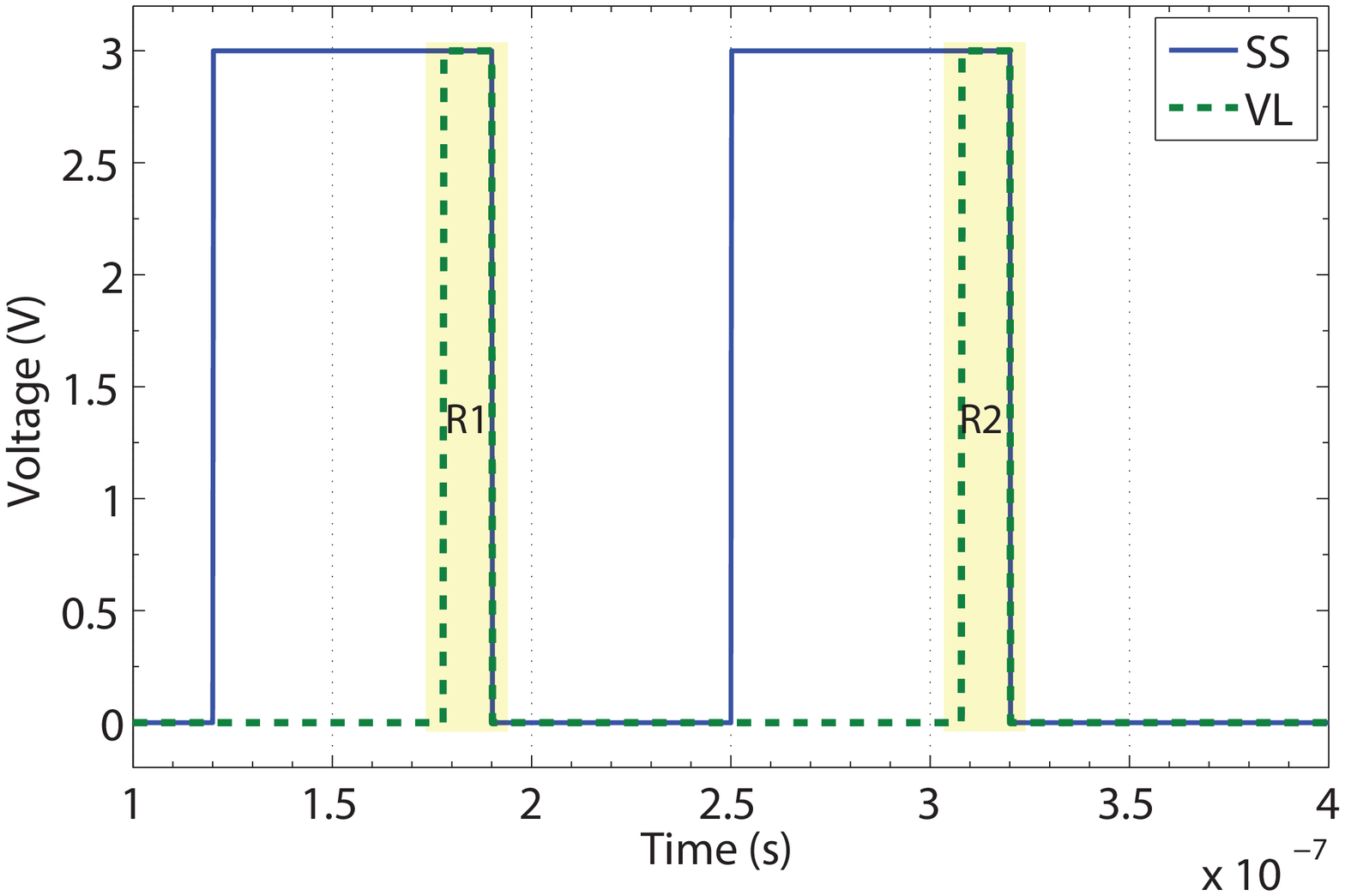,width=0.4\linewidth,clip=} \\
\footnotesize{(a)} & \footnotesize{(b)} \\ 
\epsfig{file=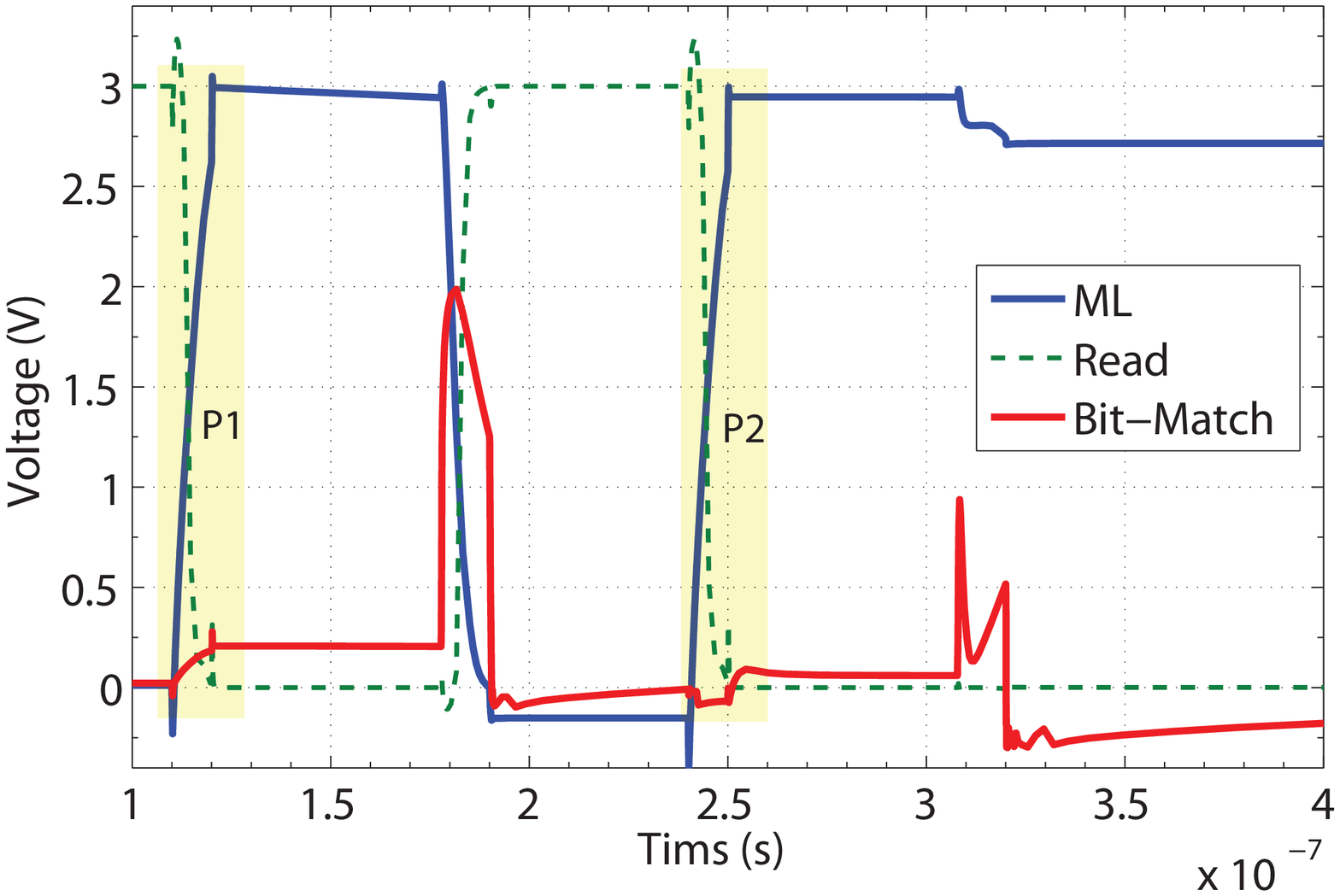,width=0.4\linewidth,clip=} &
\epsfig{file=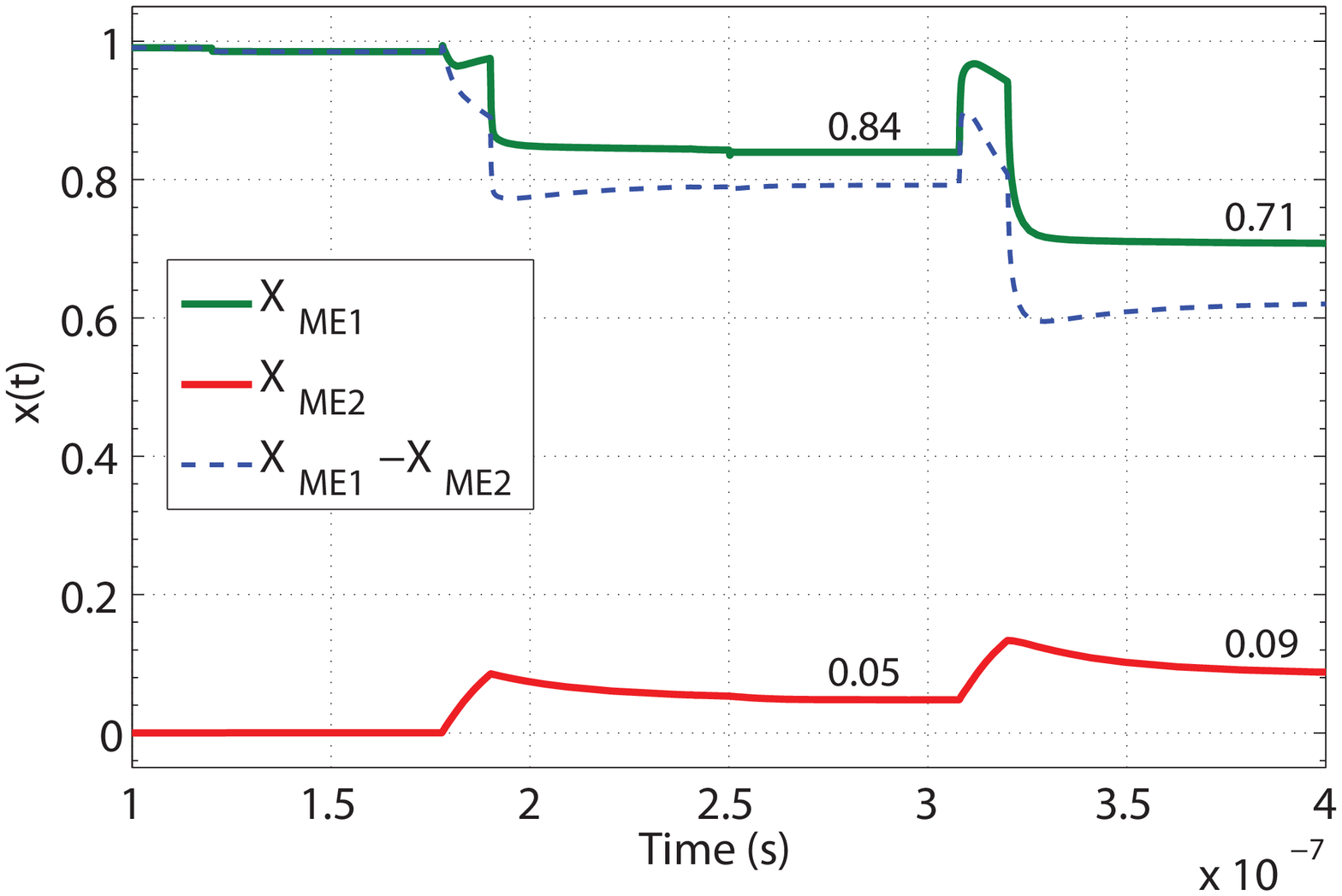,width=0.4\linewidth,clip=} \\ 
\footnotesize{(c)} & \footnotesize{(d)}
\end{tabular}

  \caption{Read operation timing diagram: (a) Search signal (S). For matching ``1'' S=$V_{\rm DD}$ and for matching ``0'' S=0, (b) Search select (SS) and read enable (VL) signals. VL$_{\rm active}=3.0$ V ($V_{\rm DD}$), (c) Bit-match, read, and match-line (ML) signals. Read=$\overline{\rm ML}$, (d)  ME1 and ME2 state variable signals. In (b) and (c), R1, R2, P1, and P2 represent two read and match-line pre-charge phases, respectively. The final (stable) values for $x_{\rm ME1}$ and $x_{\rm ME2}$ after two read operations are around $0.7$ and $0.09$. The difference between $x_{\rm ME1}$ and $x_{\rm ME2}$, in terms of time is also shown in (d).}
  \label{fig:readph}
\end{figure*}

Let us assume that ME1 and ME2 were programmed as a logic ``1'' and logic ``0'', respectively. Therefore, ME1 and ME2 are in the ``ON'' and ``OFF'' states and $R_{\rm INIT,ME1}=200~\Omega$ and $R_{\rm INIT,ME2}=99~{\rm k}\Omega$. In this case, the search line, S, is activated first. At the same time search select signal, SS, is activated to turn on the two select transistors, M5 and M6. The word select (WS) is disabled during the read operation. Fig.~\ref{fig:readph} shows the waveforms for a complete read cycle. Read operation requires higher voltage for a short period of time. The VL pulse width (PW) for read operation is $12~{\rm ns}$ as illustrated in Fig.~\ref{fig:readph}(b) which is the ``minimum'' pulse width necessary to retain memristor's state.

For a matching ``1'' (when S=$V_{\rm DD}$), the sequence of operations are as follows: (i) match line, ML, is pre-charged, (ii) SS is activated, and (iii) VL is enabled as is shown in Fig.~\ref{fig:readph}(a)-(c). A logic ``1'' is transferred to the bit-match node, which discharges the match line, ML$_i$, through transistor ML. At this point $x_{\rm ME1}$ commences to decrease its state from $1$ to $0.84$ and $x_{\rm ME2}$ increases its state from $0$ to $0.05$. Thus, there is a match between stored Data and Search Data. The following read operation for S=``0'' follows a similar pattern as shown in Fig.~\ref{fig:readph}(c). The simulation results confirm the functionality of proposed MCAM circuitry.

	\subsection{Simulation results analysis}
\label{sec:sub:compari}

Table~\ref{tab:comp} provides a comparison between the various MCAM cells that are proposed in Fig.~\ref{fig:CAMcells}. It is worth noting that simulations are based on a single cell. Therefore there are no differences in characteristics between 7-T NAND and 7-T NOR cells. The difference in minimum VL pulse width for read operation (VL$_{\rm min.PW,R}$), between different MCAM cells, is relatively significant and is brought about as the result of pass-transistors in the path from search line to the bit-match node. One important issue in the design of MCAM cells is endurance. For instance, DRAM cells must be refreshed at least every $16~{\rm ms}$, which corresponds to at least $10^{10}$ write cycles in their life cycle~\cite{Lewis2009}. Analysing a write operation followed by two serial read operations shows that 5-T, 6-T, and 7-T NOR/NAND cells deliver a promising result. After two serial read operations the memristor state values for $x_{\rm ME1}$ and $x_{\rm ME2}$ are, $0.74$ and $0.06$, and $0.71$ and $0.09$, for 5-T, 6-T, and 7-T NOR/NAND cell, respectively. The overall conclusion from the simulation results shows that in terms of speed, the 6-T NOR-type MCAM cell has improved performance, but it uses separate Data and Search lines. The 7-T NOR/NAND cell shares the same line for Data and Search inputs. However, it is slightly slower VL$_{\rm min.PW,R}=12~{\rm ns}$, while the swing on the match-line is reduced by threshold voltage ($V_{\rm th}$) drop.

\begin{table*}[hbpt]
\centering
\caption{Comparison between the proposed CAM cells in Fig.~\ref{fig:CAMcells}.}
\begin{tabular}{lcccc}
\hline
\hline
\footnotesize \multirow{2}{*}{Cell name} & 
\footnotesize VL$_{\rm min.PW,W}$ [ns] & 
\footnotesize VL$_{\rm min.PW,R}$ [ns] & 
\footnotesize V$_{\rm drop}($bit-match$)$ & 
\footnotesize Data \& Search \\ 
& 
\footnotesize VL$_{\rm W}$=$V_{\rm DD}$/2 & 
\footnotesize VL$_{\rm R}$=$V_{\rm DD}$ & 
\footnotesize Voltage [V] & 
\footnotesize Buses \\
\hline
\hline
\footnotesize  6-T NOR (Fig.~\ref{fig:CAMcells}(b)) & 
\footnotesize $223$ & 
\footnotesize $5$ & 
\footnotesize $0$ & 
\footnotesize Separate \\
\footnotesize 5-T NOR (Fig.~\ref{fig:CAMcells}(a)) & 
\footnotesize $219$ & 
\footnotesize $9$ & 
\footnotesize $V_{\rm th}$ & 
\footnotesize Separate \\
\footnotesize 7-T NOR/NAND (Fig.~\ref{fig:CAMcells}(c/d)) & 
\footnotesize $220$ & 
\footnotesize $12$ & 
\footnotesize $V_{\rm th}$ & 
\footnotesize Shared \\
\hline
\hline
\end{tabular}
\label{tab:comp}
\end{table*}

	\subsubsection{Power Analysis}
	\label{sec:sub:sub:power}

A behavioral model was used to estimate peak, average, and RMS power dissipation of an MCAM cell compared to the conventional SRAM-based cell. The power consumption is the total value for the static and dynamic power dissipation. A reduction of some $96$\% in average power consumption with an MCAM cell was noted. The maximum power dissipation reduction is over $74$\% for the memristor-based structure. The Root Mean Square (RMS) value of current, which is sunk from the supply rail for the MCAM, is around $47$~$\mu$A less than the conventional SRAM-based circuitry, which shows over $95$\% reduction. To the best of our knowledge this is the first power consumption analysis of a memristor-based structure using a behavioral modeling approach. As the technology matures it is conjectured that a similar power source could be used for the hybrid scaled CMOS/Memristor cell.


	\subsection{A $2\times 2$ Structure Verification}
	\label{sec:sub:2x2}

\begin{figure}[thpb]
\centering
\begin{tabular}{c}
\epsfig{file=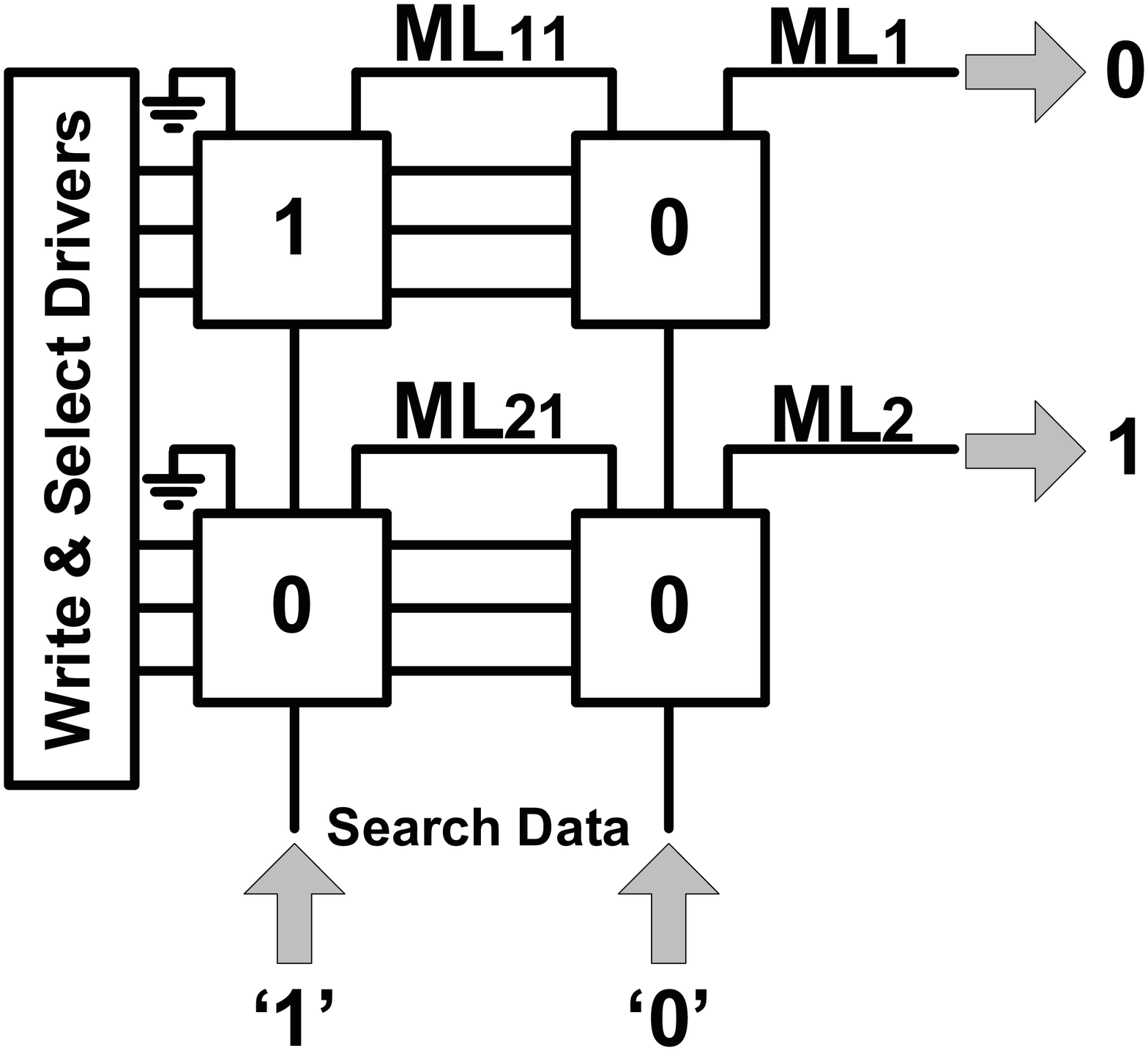,width=0.5\linewidth,clip=} \\
\footnotesize{(a) 2$\times$2 architecture, search data (``10''), and matching information} \\
\epsfig{file=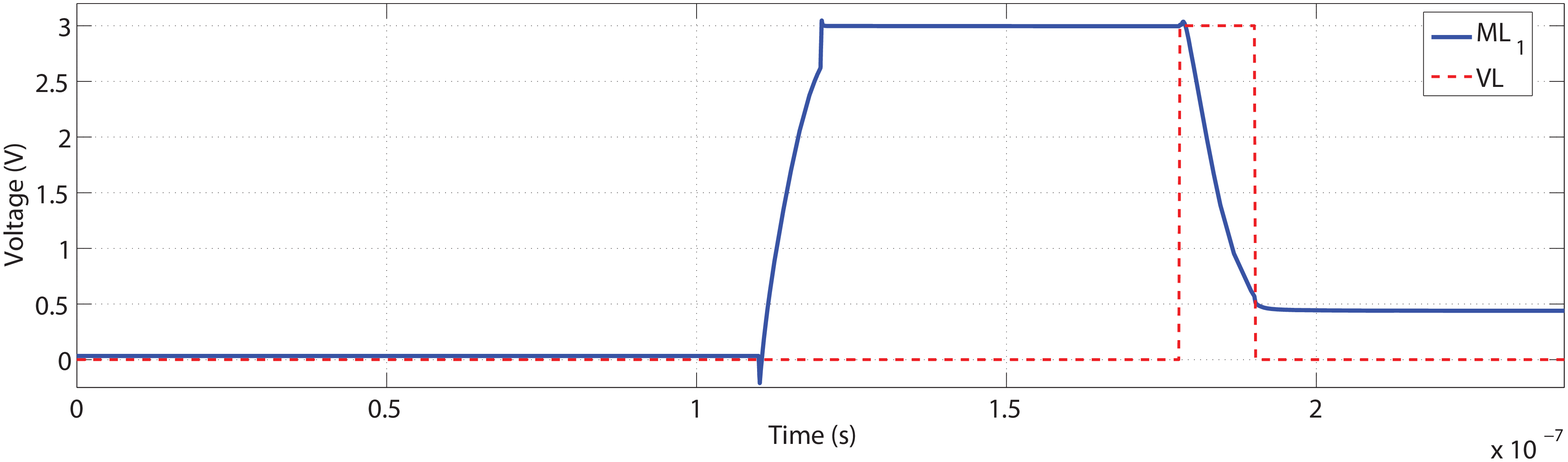,width=0.9\linewidth,clip=} \\
\footnotesize{(b) ML$_1$ signal behavior once VL triggers matching operation} \\ 
\epsfig{file=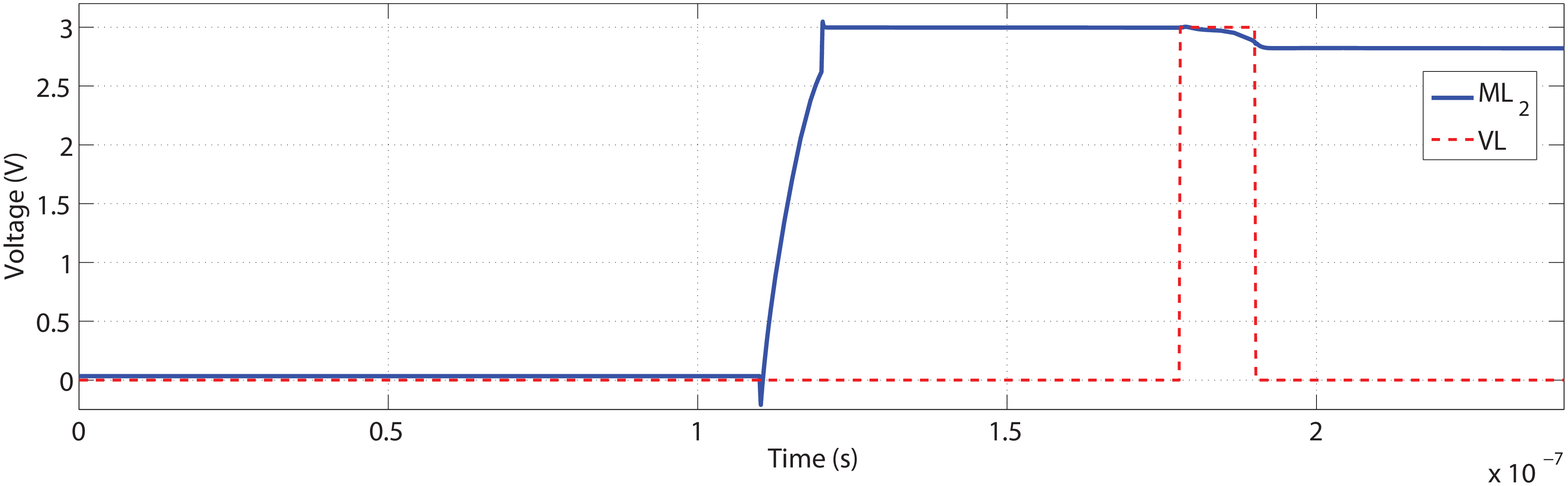,width=0.9\linewidth,clip=} \\
\footnotesize{(c) ML$_2$ signal behavior once VL triggers matching operation} 
\end{tabular}

  \caption{A 2$\times$2 MCAM structure: (a) 2$\times$2 architecture. (b) ML$_1$ signal. (c) ML$_2$ signal. The search data (``10'') is matched with the first row stored information so the ML$_1=0$ shows the search data is matched with row$_1$ and ML$_2=1$ shows the data is not matched with the stored information in the second row (row$_2$).}
  \label{fig:2x2arch}
\end{figure}

Fig.~\ref{fig:2x2arch} illustrates implementation of a $2\times 2$ structure whereby the 7-T NAND-type (Fig.~\ref{fig:CAMcells}(d)) is used. As is stated before, in the NOR-type, ML makes a connection between shared ML and ground while in the NAND-type, the ML transistors act as a series of switches between the ML$_{\rm out}$ and ground. The ML$_{1}$ and ML$_{2}$ match signals, illustrated in Fig.~\ref{fig:2x2arch}(a), are these ML$_{\rm out}$ signals. The cells are initially programmed to be ``0'' or ``1'' and the search bit vector is ``10''. The first row cells are programmed ``10''. As the consequence, ML$_1$ is discharged since there is a match between the stored and search bit vectors. Fig.~\ref{fig:2x2arch}(b) and (c) demonstrate the ML$_1$ and ML$_2$ outputs, respectively. Basically, using the ML transistors as an array of pass-transistors in a NAND-type structure imposes a significant delay, but in this case, the timing information shows the delay of matching process is around $12~{\rm ns}$.

A large scale co-simulation of crossbar memories can be carried out each junction assumed to be either a diode or a 1D-1R (a parallel structure of one diode and one resistor) or even a linear resistor~\cite{Ziegler2003}. However, the modeling approach should be carefully revisited since large resistor nonlinearity is associated with crosspoint devices~\cite{Vontobel2009}. A co-simulation of crossbar memories, considering the highly nonlinear crosspoint junctions, is underpins our longer term research objective.

	\section{Physical Layout and Fabrication}
\label{sec:layfab}


	\subsection{Physical Layout}
	\label{sec:sub:Physical-Layout-and-Layer-Definitions}
\begin{figure}[thpb]
\centering
\begin{tabular}{cc}
\epsfig{file=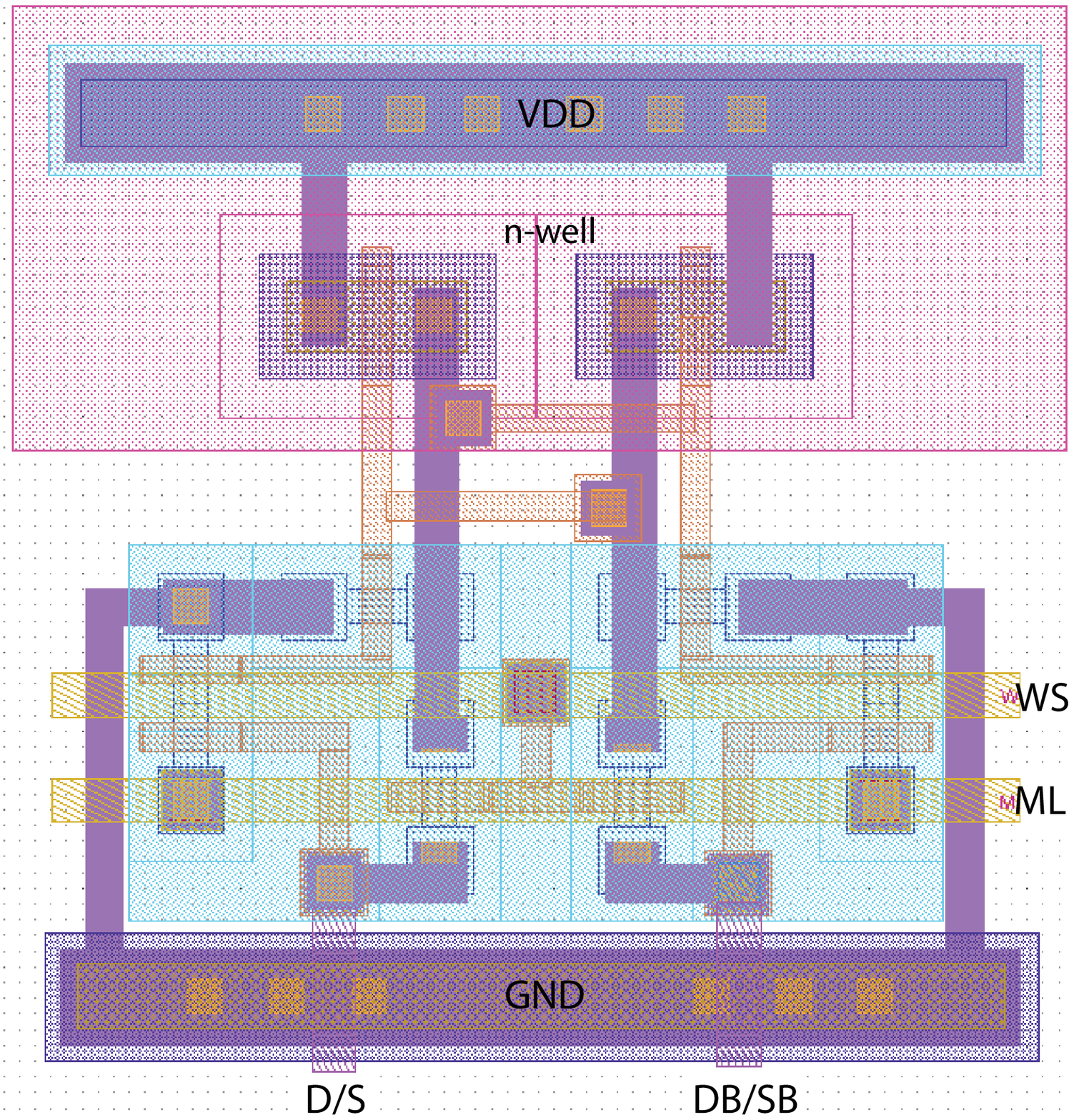,width=0.55\linewidth,clip=} \\
\footnotesize{(a) Conventional 10-T NOR-type CAM cell} \\
\epsfig{file=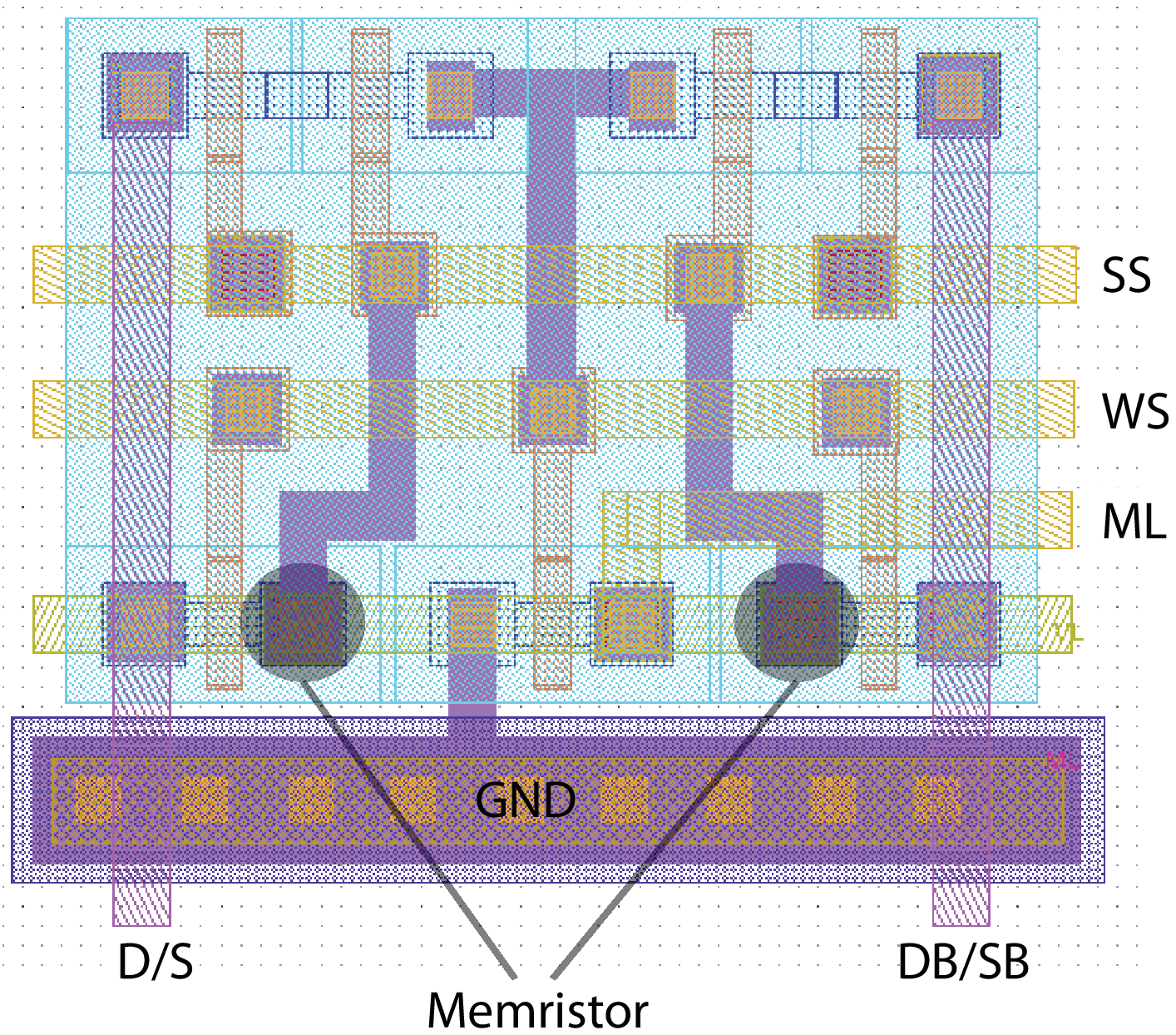,width=0.5\linewidth,clip=} \\
\footnotesize{(b) 7-T, 2-M NOR-type MCAM cell}
\end{tabular}

  \caption{Layout implementation (a) conventional SRAM-based and (b) proposed MCAM cells. In (a) $V_{\rm DD}$ line is required. In (b), highlighted regions show the two memristors in the upper layer.}
  \label{fig:layout}
\end{figure}

	Layout of conventional 10-T NOR-type CAM and 7-T NOR-type MCAM cells are shown in Fig.~\ref{fig:layout}. The MCAM cell has a dimensions of $4.8\times 4.36~\mu {\rm m}^2$ while the dimensions for the conventional SRAM-based cell is $6.0\times 6.5~\mu {\rm m}^2$. Thus, the reduction in silicon area is in the order of $46$\%. The $2\times 2$ structure also shows over a $46$\% area reduction. The two memristors, shown in highlighted regions of Fig.~\ref{fig:layout}(b) are implemented between metal-3 and metal-4 layers as part of CMOS post processing.

	\subsection{Fabrication and Layer Definitions}
	\label{sec:sub:Fabrication}
\begin{figure}[thpb]
\centering
\begin{tabular}{cc}
\epsfig{file=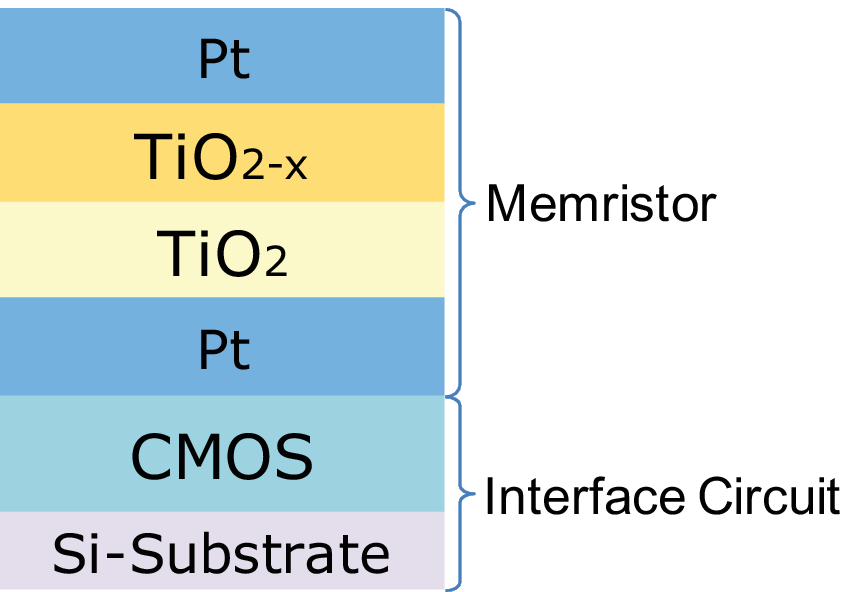,width=0.55\linewidth,clip=} \\
\footnotesize{(a) Cross section of memristor-MOS layout} \\
\epsfig{file=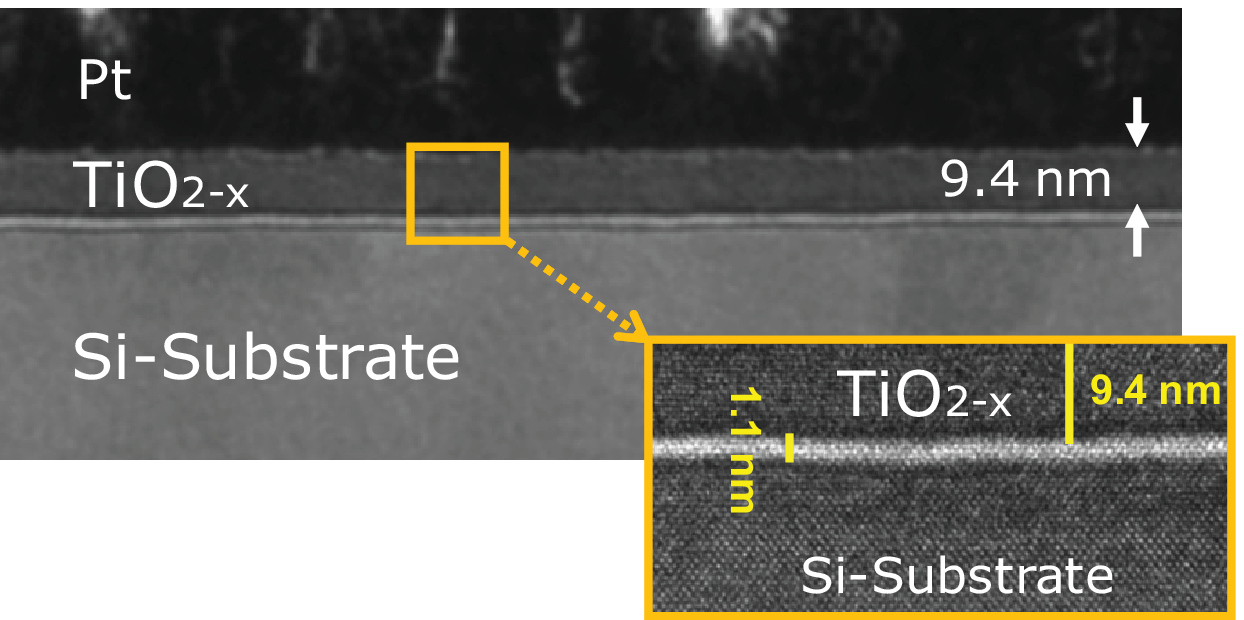,width=0.65\linewidth,clip=} \\
\footnotesize{(b) TEM microphotograph}
\end{tabular}

  \caption{A cross sectional view of the memristor-MOS implementation and TEM microphotograph of TiO$_{2-x}$ deposition.}
  \label{fig:fab}
\end{figure}

	Fig.~\ref{fig:fab}(a) illustrates a cross-section of Pt,~TiO$_2$,~and~TiO$_{2-x}$ layers over silicon substrate. The TiO$_2$ layer thickness must be restricted below two nanometers, to prevent separate conduction through the individual layers. The n-type MOS devices are patterned onto a silicon wafer using normal CMOS processing techniques, which subsequently is covered with a protective oxide layer. The Pt memristor wires are patterned and connections made to the n-type MOS devices. The upper Pt nanowire is patterned and, electrical connections made by photolithography (to spatially locate the vias) and aluminum metal deposition~\cite{Strukov2008}. 

Fig.~\ref{fig:fab}(b) demonstrates a TEM microphotograph of a TiO$_{2-x}$ overlay on a silicon substrate in order to explore the controllability of oxygen ions. The device consists of a top gate Pt, TiO$_2$/TiO$_{2-x}$ layer and back gate Pt on SiO$_2$ layer of silicon. TiO$_{2-x}$ thin film with a thickness of $9.4$~nm was deposited on a silicon wafer using sputtering technique. Table~\ref{tab:fab} is deposition result with sputtering technique. Samples show that 1.85\% oxygen (O) vacancy can be achieved keeping within the $2$\% tolerance.

\begin{table}[hbpt]
\centering
\caption{Deposition results using sputtering technique.}
\begin{tabular}{c|cccc}
\hline
\hline
\footnotesize   & 
\footnotesize O & 
\footnotesize Ti & 
\footnotesize ${\rm O}-2\times {\rm Ti}$ & 
\footnotesize $({\rm O}-2\times {\rm Ti})/{\rm Ti}$ 
\\ 
\scriptsize  & 
\scriptsize \% & 
\scriptsize \% & 
\scriptsize Normalized & 
\scriptsize Normalized  
\\
\hline
\footnotesize 1 & 
\footnotesize $66.46$ & 
\footnotesize $33.54$ & 
\footnotesize $-0.62$ & 
\footnotesize $-1.85$ 
\\
\footnotesize 2 & 
\footnotesize $66.67$ & 
\footnotesize $33.32$ & 
\footnotesize $0.03$ & 
\footnotesize $0.09$ 
\\
\hline
\hline
\end{tabular}
\label{tab:fab}
\end{table}


\section{Conclusions}
\label{sec:Conclusions}
The idea of a circuit element, which relates the charge $q$ and the magnetic flux $\phi$ realizable only at the nanoscale with the ability to remember the past history of charge flow, creates interesting approaches in future CAM-based architectures as we approach the domain of multi-technology hyperintegration where optimization of disparate technologies becomes the new challenge. The scaling of CMOS technology is challenging below $10$~nm and thus nanoscale features of the memristor can be significantly exploited. The memristor is thus a strong candidate for tera-bit memory/compare logic.

The non-volatile characteristic and nanoscale geometry of the memristor together with its compatibility with CMOS process technology increases the memory cell packing density, reduces power dissipation and provides for new approaches towards power reduction and management through disabling blocks of MCAM cells without loss of stored data. Our simulation results show that the MCAM approach provides a $45$\% reduction in silicon area when compared with the SRAM equivalent cell. The Read operation of the MCAM ranges between $5$~ns to $12$~ns, for various implementations, and is comparable with current SRAM and DRAM approaches. However the Write operation is significantly longer.

Simulation results indicate a reduction of some $96$\% in average power dissipation with the MCAM cell. The maximum power reduction is over $74$\% for the
memristor-based structure. The RMS value of current sunk from the supply rail for the MCAM is also approximately $47$~$\mu$A, which correspond to over a $95$\% reduction when compared to SRAM-based circuitry. To the best of our knowledge this is the first power consumption analysis of a memristor-based structure that has been presented using a behavioral modeling approach. As the technology is better understood and matures further improvements in performance can be expected


\section{Acknowledgement}
\label{sec:Acknowledgement}
The support provided by grant No.~R33-2008-000-1040-0 from the
World Class University (WCU) project of MEST and KOSEF through CBNU is gratefully acknowledged. The authors also note the contribution of iDataMap Pty Ltd for the initial concept and gratefully acknowledge Drs Jeong Woo Kim, Han Heung Kim, and Boung Ju Lee of NanoFab in Korea Advanced Institute Science and Technology (KAIST) for their contribution towards fabrication. 




\label{sec:References}


\bibliographystyle{unsrtnat} 
\bibliography{omid_ieeetvlsi} 

\begin{thebibliography}{29}
\providecommand{\natexlab}[1]{#1}
\providecommand{\url}[1]{\texttt{#1}}
\expandafter\ifx\csname urlstyle\endcsname\relax
  \providecommand{\doi}[1]{doi: #1}\else
  \providecommand{\doi}{doi: \begingroup \urlstyle{rm}\Url}\fi

\bibitem[Bourianoff et~al.(2007)Bourianoff, Gargini, and
  Nikonov]{Bourianoff2007}
G.~I. Bourianoff, P.~A. Gargini, and D.~E. Nikonov.
\newblock Research directions in beyond {CMOS} computing.
\newblock \emph{Solid-State Electronics}, 51\penalty0 (11-12):\penalty0 1426 --
  1431, 2007.

\bibitem[Akinwande et~al.(2008)Akinwande, Yasuda, Paul, Fujita, Close, and
  Wong]{Akinwande2008}
D.~Akinwande, S.~Yasuda, B.~Paul, S.~Fujita, G.~Close, and H.~S.~P. Wong.
\newblock Monolithic integration of {CMOS} {VLSI} and {CNT} for hybrid
  nanotechnology applications.
\newblock In \emph{Proc. 38th European Solid-State Device Research Conference,
  ESSDERC'08}, pages 91--94, 2008.

\bibitem[Engheta(2007)]{Engheta2007}
N.~Engheta.
\newblock Circuits with light at nanoscales: Optical nanocircuits inspiredby
  metamaterials.
\newblock \emph{Science}, 317\penalty0 (5845):\penalty0 1698--1702, 2007.

\bibitem[Strukov et~al.(2008)Strukov, Snider, Stewart, and
  Williams]{Strukov2008}
D.~B. Strukov, G.~S. Snider, D.~R. Stewart, and R.~S. Williams.
\newblock The missing memristor found.
\newblock \emph{Nature}, 453\penalty0 (7191):\penalty0 80--83, 2008.

\bibitem[Chua(1971)]{Chua1971}
L.~O. Chua.
\newblock Memristor - the missing circuit element.
\newblock \emph{IEEE Transactions on Circuits and Systems}, 18\penalty0
  (5):\penalty0 507--519, 1971.

\bibitem[Kang(1975)]{Kang1975}
S.~M. Kang.
\newblock \emph{On The Modeling of Some Classes of Nonlinear Devices and
  Systems}.
\newblock {Doctoral dissertation in Electrical and Electronics Engineering},
  76-15-251, {University of California, Berkeley, CA}, 1975.

\bibitem[Chua and Kang(1976)]{Chua1976}
L.~O. Chua and S.~M. Kang.
\newblock Memristive devices and systems.
\newblock \emph{Proceedings of the IEEE}, 64\penalty0 (2):\penalty0 209--223,
  1976.

\bibitem[Kavehei et~al.(2009)Kavehei, Kim, Iqbal, Eshraghian, Al-Sarawi, and
  Abbott]{Kavehei2009}
O.~Kavehei, Y.~S. Kim, A.~Iqbal, K.~Eshraghian, S.~F. Al-Sarawi, and D.~Abbott.
\newblock The fourth element: Insights into the memristor.
\newblock In \emph{the IEEE International Conference on Communications,
  Circuits and Systems, ICCCAS}, pages 921--927, July 2009.

\bibitem[Yang et~al.(2008)Yang, Pickett, Li, Ohlberg, Stewart, and
  Williams]{Yang2008}
J.~Joshua Yang, Matthew~D. Pickett, Xuema Li, Douglas A.~A. Ohlberg, Duncan~R.
  Stewart, and R.~Stanley Williams.
\newblock Memristive switching mechanism for metal-oxide-metal nanodevices.
\newblock \emph{Nature Nanotechnology}, 3\penalty0 (7):\penalty0 429--433,
  2008.

\bibitem[Strukov and Williams(2009{\natexlab{a}})]{Strukov2009}
D.~B. Strukov and R.~S. Williams.
\newblock Exponential ionic drift: Fast switching and low volatility of
  thin-film memristors.
\newblock \emph{Applied Physics A: Materials Science and Processing},
  94\penalty0 (3):\penalty0 515--519, 2009{\natexlab{a}}.

\bibitem[Biolek et~al.(2009)Biolek, Biolek, and Biolkov{\'a}]{Biolek2009}
Z.~Biolek, D.~Biolek, and V.~Biolkov{\'a}.
\newblock {SPICE} model of memristor with nonlinear dopant drift.
\newblock \emph{Radioengineering Journal}, 18\penalty0 (2):\penalty0 211, 2009.

\bibitem[Benderli and Wey(2009)]{Benderli2009}
S.~Benderli and T.~A. Wey.
\newblock On {SPICE} macromodelling of {TiO$_2$} memristors.
\newblock \emph{Electronics Letters}, 45\penalty0 (7):\penalty0 377--379, 2009.

\bibitem[Joglekar and Wolf(2009)]{Joglekar2009}
Y.~N. Joglekar and S.~J. Wolf.
\newblock The elusive memristor: Properties of basic electrical circuits.
\newblock \emph{European Journal of Physics}, 30\penalty0 (4):\penalty0 661,
  2009.

\bibitem[Shin et~al.(2010)Shin, Kim, and Kang]{Shin2010}
S.~Shin, K.~Kim, and S.~M. Kang.
\newblock Compact models for memristors based on charge-flux constitutive
  relationships.
\newblock \emph{IEEE Transactions on Computer-Aided Design of Integrated
  Circuits and Systems}, 29\penalty0 (4):\penalty0 590 --598, April 2010.

\bibitem[{International Technology Roadmap for Semiconductors}()]{ITRS2009}
{International Technology Roadmap for Semiconductors}.
\newblock {Emerging Research Devices (ERD) 2009 Edition}.
\newblock \url{http://www.itrs.net/}.

\bibitem[Freitas and Wilcke(2008)]{Freitas2008}
R.~F. Freitas and W.~W. Wilcke.
\newblock Storage-class memory: The next storage system technology.
\newblock \emph{IBM Journal of Research and Development}, 52\penalty0
  (4-5):\penalty0 439--448, 2008.

\bibitem[Kuekes et~al.(2005)Kuekes, Stewart, and Williams]{Kuekes2005}
P.~J. Kuekes, D.~R. Stewart, and R.~S. Williams.
\newblock The crossbar latch: Logic value storage, restoration, and inversion
  in crossbar circuits.
\newblock \emph{Journal of Applied Physics}, 97:\penalty0 034301, 2005.

\bibitem[Strukov and Williams(2009{\natexlab{b}})]{Strukov2009b}
D.~B. Strukov and R.~S. Williams.
\newblock Four-dimensional address topology for circuits with stacked
  multilayer crossbar arrays.
\newblock \emph{Proceedings of the National Academy of Sciences}, 106\penalty0
  (48):\penalty0 20155--20158, 2009{\natexlab{b}}.

\bibitem[Vontobel et~al.(2009)Vontobel, Robinett, Kuekes, Stewart, Straznicky,
  and Williams]{Vontobel2009}
P.~O. Vontobel, W.~Robinett, P.~J. Kuekes, D.~R. Stewart, J.~Straznicky, and
  R.~S. Williams.
\newblock Writing to and reading from a nano-scale crossbar memory based on
  memristors.
\newblock \emph{Nanotechnology}, 20\penalty0 (42):\penalty0 425204, 2009.

\bibitem[Tyshchenko and Sheikholeslami(2008)]{Tyshchenko2008}
O.~Tyshchenko and A.~Sheikholeslami.
\newblock Match sensing using match-line stability in content-addressable
  memories {(CAM)}.
\newblock \emph{IEEE J Solid-St Circ}, 43\penalty0 (9):\penalty0 1972--1981,
  2008.

\bibitem[Kumaki et~al.(2007)Kumaki, Kuroda, Ishizaki, Koide, Mattausch, Noda,
  Dosaka, Arimoto, and Saito]{Kumaki2007}
T.~Kumaki, Y.~Kuroda, M.~Ishizaki, T.~Koide, H.~J. Mattausch, H.~Noda,
  K.~Dosaka, K.~Arimoto, and K.~Saito.
\newblock Real-time {Huffman} encoder with pipelined {CAM}-based data path and
  code-word-table optimizer.
\newblock \emph{IEICE - Transactions on Information and Systems},
  E90-D\penalty0 (1):\penalty0 334--345, 2007.

\bibitem[Kim et~al.(2009)Kim, Ahn, Kim, and Jeong]{Kim2009b}
Y.~D. Kim, H.~S. Ahn, S.~Kim, and D.~K. Jeong.
\newblock {A High-Speed Range-Matching TCAM for Storage-Efficient Packet
  Classification}.
\newblock \emph{IEEE Transactions on Circuits and Systems}, 56\penalty0
  (6):\penalty0 1221--1230, June 2009.

\bibitem[Verma and Chandrakasan(2008)]{Verma2008}
N.~Verma and A.P. Chandrakasan.
\newblock A 256~kb$\times$65~nm 8{T} subthreshold {SRAM} employing
  sense-amplifier redundancy.
\newblock \emph{IEEE J Solid-St Circ}, 43\penalty0 (1):\penalty0 141--149, Jan.
  2008.

\bibitem[Lu and Hsu(2006)]{Lu2006}
S.~K. Lu and C.~H. Hsu.
\newblock {Fault Tolerance Techniques for High Capacity RAM}.
\newblock \emph{IEEE Transactions on Reliability}, 55\penalty0 (2):\penalty0
  293--306, June 2006.

\bibitem[Mohan and Sachdev(2009)]{Mohan2009}
N.~Mohan and M.~Sachdev.
\newblock Low-leakage storage cells for ternary content addressable memories.
\newblock \emph{IEEE Transactions on Very Large Scale Integration (VLSI)
  Systems}, 17\penalty0 (5):\penalty0 604--612, May 2009.

\bibitem[Pagiamtzis and Sheikholeslami(2006)]{Pagiamtzis2006}
K.~Pagiamtzis and A.~Sheikholeslami.
\newblock Content-addressable memory (cam) circuits and architectures: A
  tutorial and survey.
\newblock \emph{IEEE Journal of Solid-State Circuits}, 41\penalty0
  (3):\penalty0 712--727, March 2006.

\bibitem[Witrisal(2009)]{Witrisal2009a}
K.~Witrisal.
\newblock Memristor-based stored-reference receiver - the {UWB} solution?
\newblock \emph{Electronics Letters}, 45\penalty0 (14):\penalty0 713--714,
  2009.

\bibitem[Lewis and Lee(2009)]{Lewis2009}
D.~L. Lewis and H.~H.~S. Lee.
\newblock Architectural evaluation of {3D} stacked {RRAM} caches.
\newblock In \emph{{IEEE International Conference on 3D System Integration,
  3DIC}}, pages 1--4, San Francisco, CA, 2009.

\bibitem[Ziegler and Stan(2003)]{Ziegler2003}
M.~M. Ziegler and M.~R. Stan.
\newblock Cmos/nano co-design for crossbar-based molecular electronic systems.
\newblock \emph{IEEE Transactions on Nanotechnology}, 2\penalty0 (4):\penalty0
  217--230, 2003.

\end{thebibliography}

%








\end{document}